\documentclass[journal]{IEEEtran}
\usepackage{cite}

\ifCLASSINFOpdf
 \usepackage[pdftex]{graphicx}
\else
\fi
\usepackage{amsmath}
\usepackage{algorithmic}
\ifCLASSOPTIONcompsoc
\usepackage[caption=false,font=normalsize,labelfont=sf,textfont=sf]{subfig}
\else
\usepackage[caption=false,font=footnotesize]{subfig}
\fi

\usepackage{color}
\usepackage{amssymb}	
\newcommand{\bfb}{\mathbf{b}}
\newcommand{\bfu}{\mathbf{u}}
\renewcommand{\Re}{\mbox{Re}}
\renewcommand{\Im}{\mbox{Im}}

\begin{document}
%
\title{A New Algorithm for Improved Blind Detection of Polar Coded PDCCH in 5G New Radio}
%
%
%

\author{Amin~Jalali,~\IEEEmembership{Student Member,~IEEE,}
        and~Zhi~Ding,~\IEEEmembership{Fellow,~IEEE}

\thanks{The authors are with the Department of Electrical and Computer Engineering,
University of California at Davis, Davis, CA 95616 USA (e-mail:
amjal@ucdavis.edu; zding@ucdavis.edu).}
\thanks{This work was presented in part at the 2018 IEEE International Symposium on Information Theory in Vail, Colorado, USA}
\thanks{This material is based on works supported by the National
Science Foundation under Grants 1700762 and 1711823}}

%
%


\maketitle

\begin{abstract}

In recent release of the new cellular standard known as 5G New Radio (5G-NR),
the physical downlink control channel (PDCCH) has adopted polar codes for 
error protection. Similar to 4G-LTE, each active user equipment (UE) must blindly detect its own PDCCH in the downlink search space. 
This work investigates new ways to improve the accuracy of PDCCH
blind detection in 5G-NR.  
We develop a novel design of joint detection and decoding receiver 
for 5G multiple-input multiple-output (MIMO) transceivers. We aim to
achieve robustness against practical obstacles including channel state 
information (CSI) errors, noise, co-channel interferences, and pilot contamination. 
To optimize the overall receiver performance in PDCCH blind detection, we 
incorporate the polar code information during the signal detection stage
by relaxing and transforming the Galois field code constraints into the
complex signal field. Specifically,
we develop a novel joint linear programming (LP) formulation 
that takes into consideration the transformed polar code constraints.
Our proposed joint LP formulation can also be integrated with polar decoders
to deliver superior receiver performance
at low cost.
We further introduce a metric that can be used to eliminate most of wrong 
PDCCH candidates to improve the computational efficiency
of PDCCH blind detection for 5G-NR. 
\end{abstract}


%
\IEEEpeerreviewmaketitle

\section{Introduction}
%
%
%
%
\IEEEPARstart{I}{n} modern mobile wireless communication standards such as the 
3GPP LTE/LTE-Advanced \cite{TS36.213}
and the 3GPP NR\cite{TS38.213}, active user-equipment (UE) devices must obtain crucial control 
messages known as downlink control information (DCI) from the serving gNB.
It is highly critical for the UE to correctly receive its DCI that are placed
in the PDCCH search space since receiving data which will be sent in physical 
downlink shared channel (PDSCH) depends on it. 
Since a UE does not know the location of its DCI in the search space, 
it must perform so-called blind detection by attempting to decode a set of possible 
DCI candidates.
According to the 3GPP standards \cite{TS36.213}\cite{TS38.213},
the number of such candidates may even exceed forty.  
Thus, using 
a complex decoder for error-correcting codes on all candidates is costly 
in terms of computation, latency, and power consumption. Therefore, 
it is of practical importance to investigate means for
improving both the detection performance and the computation efficiency.

In the 4G cellular communication standards by the 3GPP, the control message (DCI) 
is encoded with convolutional codes for which
blind detection algorithms have been designed \cite{shieh2005strategies,moosavi2011fast,sipila2012blind,malladi2012methods}. 
Other blind-detection algorithms for low-density parity check (LDPC) codes \cite{xia2014novel} 
and Bose-Chaudhuri-Hocquenghem (BCH) \cite{zhou2013information} codes have also been proposed. 
However, in the 5G cellular communication standard (5G), a more interesting
class of forward error correcting (FEC) codes known as polar codes 
are adopted for error protection of PDCCH. 
Polar codes, discovered by Ar{\i}kan in 2008, have been shown to achieve channel capacity on 
binary-input discrete memoryless channels (BDMCs) without high encoding and decoding 
complexities \cite{arikan2009channel, arikan2008performance}. Many investigative works on polar codes 
have appeared, including their construction and use under additive white Gaussian noise (AWGN) channels \cite{arikan2008performance, tal2013construct, wu2014construction} and fading channels 
\cite{bravo2013polar}. Successful and popular  decoder algorithms include successive 
cancellation (SC) decoding \cite{arikan2009channel} 
and successive cancellation 
list (SCL) decoding introduced in \cite{tal2011list}. 
Effective algorithms for the blind detection of
polar coded DCI in 5G cellular networks remain an open problem. 
Accurate detection of polar coded DCI by the UE is 
of critical importance to the success of cellular communications. 
This work focuses the development of reliable and low complexity 
blind detection of DCI signals by jointly exploiting the 
antenna diversity and polar code constraints to develop a novel and integrative
receiver algorithm. 

In most modern wireless communication systems, multiple-input multiple-output (MIMO) transceivers 
have been widely adopted owing to their ability to achieve 
high spectral efficiency. As a result, robust and efficient receivers for MIMO 
wireless systems have been widely investigated in the literature. 
It is well known that  maximum likelihood detector (MLD) can
optimally minimize the probability of detection error. Well known
MLD algorithms for MIMO systems include the sphere decoding \cite{viterbo1999universal} and a
V-BLAST detection \cite{wolniansky1998v}. 
Ideally, optimum receiver performance can be accomplished 
by applying the principle of maximum likelihood
detection under the FEC codeword constraints to reject that symbol sequences that 
belong to invalid FEC codewords. 
However, for most FEC codes that are not very short, joint maximum likelihood detection
and decoding (MLDD) can be extraordinarily complex and costly to implement.
As a result, most practical MIMO receivers apply MIMO detection 
followed by FEC decoding separately. The lack of joint MLDD
is attributed to the NP-hard high complexity of incorporating Galois field FEC constraints 
into maximum likelihood symbol detection criteria that operate strictly in 
Euclidean (real or complex) field.
To overcome the difficulty posed by the conflicting fields of detection versus decoding, 
existing joint MIMO detection and FEC decoding 
receivers typically utilize the exchange of soft information between soft
decoder and detector to form a turbo receiver 
\cite{douillard1995iterative, lu2002ldpc, hochwald2003achieving, wang2018joint, wang2018integrated}.

Recently, a new type of receiver based on joint detection and decoding 
has been proposed \cite{wang2015joint, wang2014joint}.  
Transforming the Galois field code constraints into 
a set of linear inequality constraints in the Euclidean field,
a joint linear programing (LP) receiver can leverage the constraints
imposed by modulated data symbols, training symbols, noise subspace,
and the LPDC code to achieve improved performance. The concept of joint 
detection-decoding has been applied in massive MIMO systems to combat 
pilot contamination \cite{wang2016robust}\cite{wang2016fec} and also used to 
tackle channel uncertainties such as partial channel information 
\cite{wang2015diversity,wang2016diversity} or imperfect channel estimation \cite{jalali2018joint}.

In this work, we study how the joint detection-decoding principle based on
integrated constraints in Euclidean field can improve the receiver performance
in the practical PDCCH detection for 5G cellular receivers. 
Our chief objective is to enable UE receivers to 
eliminate false PDCCH candidates during the signal detection
stage and to reduce receiver delay and energy consumption. 
We also aim to develop UE receivers to be
reliable and robust against practical non-idealities
including CSI errors, noises,
co-channel interferences, and pilot contamination. 
To achieve our goals, we formulate an LP optimization
problem for the joint MIMO detector that
incorporates relaxed polytope polar code constraints. 
This relaxed polytopes, proposed in \cite{goela2010lp}
are constructed according to the factor graph 
representation of polar codes and achieve the 
ML-certificate property. 
Furthermore, we take advantage of theses code constraints in the LP formulation 
in order to define a metric that we call \emph{fractional metric}. 
We show that this metric is capable of identifying the polar codewords in detection stage. 
Based on the fractional metric, we propose a new
algorithm to improve the blind detection of DCI
by eliminating all but one of the candidates for decoding. 
Substantially reducing the number of decoding steps,
our proposed blind detection algorithm is capable of saving receiver latency and
power consumption substantially while maintaining a very low probability of 
missed detection in the search space.

We note that blind detection of polar coded messages has been investigated 
recently in \cite{condo2017blind, giard2017blind} as well. 
In particular, the authors of \cite{giard2017blind} 
proposed a different detection metric for identifying whether
a received block may be a candidate block encoded by a particular 
polar code. The authors of \cite{condo2017blind} developed a blind detection 
scheme that requires transmitting the radio network temporary identifier (RNTI) 
on some of the frozen bits of the polar code before letting 
the decoder to eliminate a subset of candidates in the search space. 
On the other hand,
our proposed blind detection algorithm utilizes the code constraints to form
a novel integrative detector. Since the metric that we propose in this paper is 
readily accessible at the detector, it can be easily integrated with
the existing schemes of \cite{condo2017blind, giard2017blind}.

In this paper, we first introduce the problem of DCI blind detection in PDCCH search
space of 5G cellular networks in Section II.  We shall introduce the basic principles of
polar codes. 
In Section III,  we present a problem formulation of FEC coded MIMO detector as linear program
by incorporating the polar code constraints based on the approach in \cite{goela2010lp} to develop a novel joint LP receiver.  This new joint receiver effectively 
exploits the rich and important diversity of FEC code constraints and is
robust against practical non-idealities including CSI errors, noises,
co-channel interferences, and pilot contamination. 
Our proposed joint LP receiver can also be 
integrated with 
popular polar decoders such as SC and SCL decoders for
reduced complexity implementation. 
In Section IV, we propose a metric obtained in solving LP optimization problem 
and present an algorithm using this metric to identify the right DCI candidate in the 
search space that contain the UE control message. 
We present simulation results in Section V to illustrate the
efficacy of the proposed receiver design both in terms of BER/BLER and also its capability to improve blind detection process, before concluding this paper in section VI.

\section{System Model}
\subsection{5G Downlink Physical Parameters}
\begin{figure}
 \begin{center}
    \includegraphics[scale = 0.37]{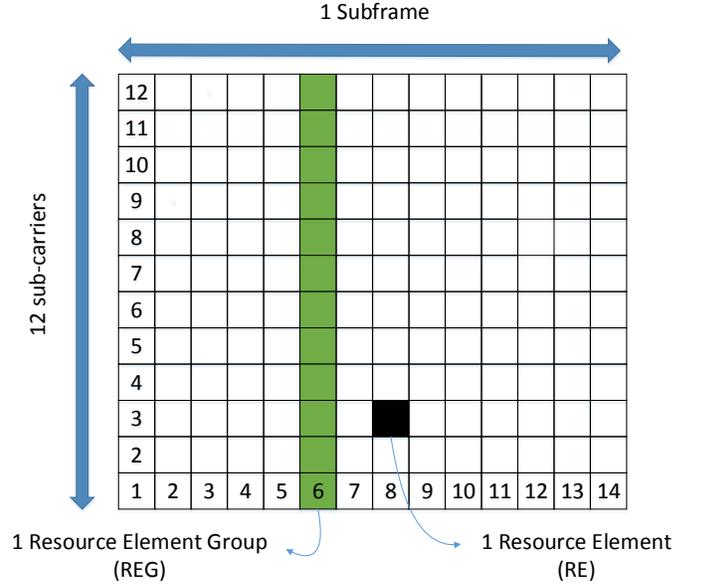}
    \caption{5G/NR frame structure terminology, Normal CP, $\mu$ = 0 (15Khz subcarrier spacing)}
     \label{fig:lte}
\end{center}
\end{figure} 
As described in 3GPP new radio (NR) standard \cite{TS38.213}, the
time intervals are expressed with a basic time unit $T_c = 1/1966080000$ seconds.
Similar to 4G-LTE, each radio frame has a length of 10 ms and 
is equally divided into ten subframes of 1 ms in duration; however, in NR, OFDM subcarrier spacing is scalable based on the numerology ($\mu$).
For $\mu = 0$ and normal cyclic prefix case, each subframe consists of 14 OFDM symbols, as illustrated in Fig. \ref{fig:lte}.

A resource element (RE) is the smallest unit of the resource grid made up 
of one OFDM sub-carrier for one OFDM symbol interval. 
A physical resource block (RB) is the smallest unit of resources 
that can be allocated to a user. Without loss of generality, we can
consider numerology $\mu=0$ \cite{TS38.213}, for which the subcarrier spacing is 15 
kHz. 
The resource block is 180 kHz wide in frequency. A resource element 
group 
(REG) is made up of 12 resource elements in frequency domain and
one OFDM symbol in time domain. Moreover, multiple continuous REGs 
constitute a control channel element (CCE) that is used to carry downlink 
control information (DCI). Aggregation level indicates how many CCEs are allocated for PDCCH, depending on the DCI size. 
Based on NR standard \cite{TS38.213}, aggregation level can be 1/2/4/8/16 . Control resource set
(CORESET), consists of multiples 
of 12 REs in frequency domain, across 1/2/3 OFDM symbols 
in time domain. CORESET is equivalent to the control region in 
LTE subframe \cite{TS38.213}.

%



The DCI uses QPSK modulation
and carries information about downlink shared channel (DL-SCH) resource
allocation. gNB schedules the downlink shared channel (DL-SCH) 
resource blocks for different users through the DCI. 
DCI also contains the 
information regarding DL-SCH hybrid automatic repeat request (HARQ). 
Clearly, it is vital for each UE to correctly
decode its DCI to receive actual downlink data messages.

\subsection{Blind Detection}
UE needs to calculate the number of available CCEs used for downlink control 
information and index them. This depends on depend on CORESET, bandwidth of the system and number of antenna ports 
present which in turn will affect the number of reference signals 
present. To calculate the number of CCEs available for PDCCH,
UE first needs to calculate the number of REs used for PDCCH by 
subtracting REs used for reference symbols, PCFICH, and PHICH, respectively, from 
the total REs that are allocated for control region. 
Suppose $3$ first symbols are allocated for control region while
the bandwidth is $10$ Mhz to allow $50$ RBs.  
Therefore $$\text{Total REs} = 3 \times 12 \times 50 = 1800 \text{REs}$$
Then we calculate number of REs for PDCCH by 

REs for PDCCH = Total REs $-$\\
  \hspace*{1.5cm} Number of REs used for reference signals $-$\\
  \hspace*{1.5cm} Number of REs used in PHICH $-$\\
  \hspace*{1.5cm} Number of REs used in PCFICH \\
$$\text{CCEs available for PDCCH} = \frac{\text{REs for PDCCH}}{\text{Number of REG per CCE} \times {12}}$$\\

Next, each active and searching UE shall group the remaining REs into CCEs and 
index them.
The first 16 CCEs belong to common search space (CSS) in which 
control information is relevant to all the receiving UEs. 
If there are more than 16 CCEs available, the remainder would be used 
to send UE-specific control information that is only decodable 
for a particular UE. This particular space is called UE specific search space (UESS).

Suppose, there are total number of $40$ CCEs that gNB can send DCI for a particular UE. 
If that particular UE does not know at which index it needs to start scanning and also how 
many consecutive CCEs consisting its DCI (aggregation level), then it would face overwhelming 
number of candidates to check. To limit the search space so as to save energy and 
computation cost, the gNB fixes some CCE indices for a particular UE. This means that
the UE knows, for each aggregation level, what CCE indices to start looking for among all 
indices, based on a formula related to its RNTI and the subframe number 
which are known to both the UE and serving gNB. 

\begin{figure}
\centering 
    \includegraphics[scale=0.32]{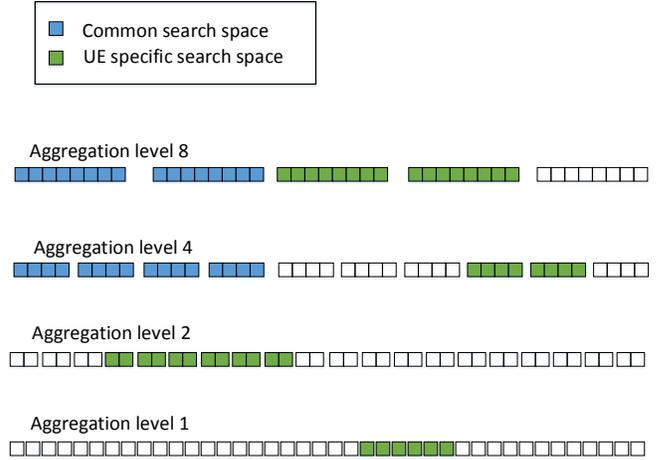}
    \caption{Common and UE specific search space for each aggregation level}
     \label{fig:fig-1}

\end{figure}

In Table \ref{table:0}, the number of CCE indices and PDCCH candidates searched by a UE in a 
subframe for each search space has been mentioned for LTE standard \cite{TS36.213}.
Figure \ref{fig:fig-1} is based on Table \ref{table:0} 
and graphically shows the hypothetical PDCCH indices to scan for each aggregation level, 
from all the potential indices. 

 \begin{table}
  \centering
\caption { Number of PDCCH candidates for different aggregation levels and search space type }
\label{table:0}
 \begin{tabular}{ |p{1.7cm}|p{1.7cm}|p{1.7cm}|p{2.2cm}|  }
\hline
search space type& aggregation level &size (in CCEs) & number of PDCCH candidates\\
 \hline
 
                        &   1     & 6   &   6\\
   UESS           &   2     & 12 &   6\\
                        &   4     &  8  &   2\\
                        &   8     &  16   &  2\\
  \hline
                        &   4     &  16  &   4\\
   CSS                     &   8     &  16   &  2\\

  \hline
\end{tabular}
\end{table}

For instance, for CSS with 16 CCEs, aggregation level can be either 4 or 8. 
There are 2 possible DCI formats in CSS. 
Therefore, in case of aggregation level 4, UE needs to 
consider CCE index 0, 4, 8 and 12. 
For aggregation level 8, it needs to consider CCE index 
0 and 8. Hence, in this case the number of PDCCH candidates 
can be calculated as:
\begin{align*}
&(4 [\text{in aggr. level } 4] + 2 [\text{in aggr. level } 8] )\\
& \times  (\text{ 2 [DCI formats in CSS]})= 12.
\end{align*}

Therefore, UE needs to calculate the CCE indices possible for each aggregation level for both CSS and UESS, to get the PDCCH candidates. After decoding each PDCCH candidate,
the UE shall use the 16-bit cyclic redundancy check (CRC) code appended to the DCI payload
in order to verify whether the candidate is the correct one for the UE. The gNB masks the
CRC using the UE specific RNTI as a distinct identifier. The
UE validates the decoded candidate by checking the unmasked CRC code. 
If a PDCCH candidate passes the CRC after RNTI unmasking, the UE 
shall claim the PDCCH candidate as its own correctly decoded DCI. 
This necessary UE process in cellular communications
is called blind detection or blind decoding in the 3GPP standards of 4G-LTE and 5G-NR.

Blind detection saves downlink bandwidth since no additional signaling is needed to indicate to the UE the location of its DCI. By letting the
UE detect its DCI in the search space, blind detection trades 
computation for bandwidth. There naturally exists accompanied latency 
and energy consumption, particularly  when a computationally expensive channel decoder is used. Thus, it is highly desirable to improve
the performance of the detection so as to reduce the set of 
candidates that are sent to the UE decoder.
Our novel proposal of a joint DCI detector 
is capable of generating a metric that helps the UE to considerably reduce the number of such search candidates. 

\subsection{Polar Code for Error Protection}
\begin{figure}
\centering 
    \includegraphics[scale=0.32]{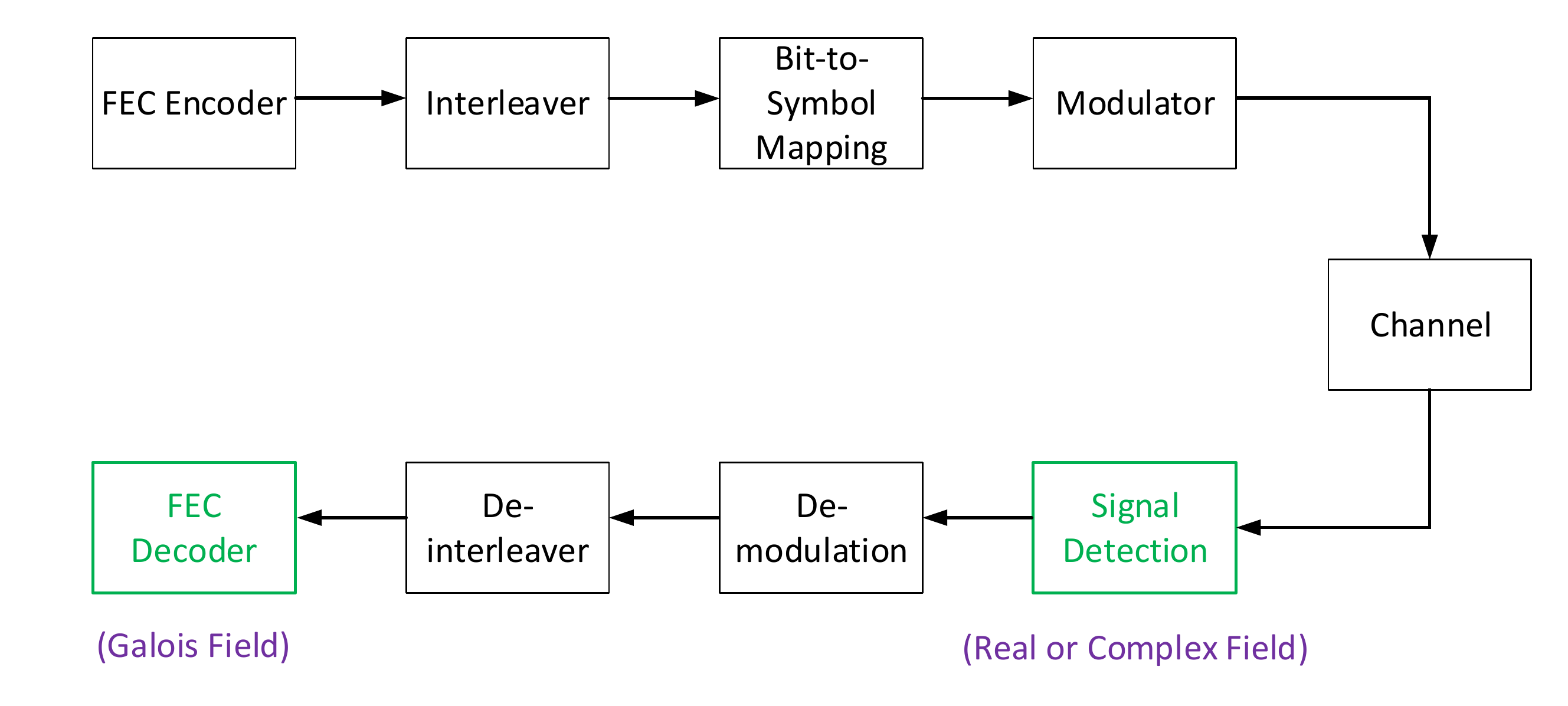}
    \caption{Polar-coded MIMO system}
     \label{fig:fig1}

\end{figure} 

Figure \ref{fig:fig1} illustrates a generic
polar-coded MIMO transmission system in which
information bits are first encoded
with polar encoder as FEC (forward error correction) codewords 
before being mapped into QAM data symbols of 
constellation $\tilde{Q}$. Note that PDCCH uses QPSK modulation
for data symbols. 
The MIMO receiver aims to recover the source bits despite 
channel distortions, interferences, and channel noises. 
Our goal is to design a joint detector that incorporates
the polar code information that is in Galois field in 
MIMO signal detection for improved performance. 

Briefly, 
a polar code of rate $R = K / N$ is specified by 
$(N, K, \mathcal{I}^c)$, where $N = 2^n$ is the codeword 
length and $K$ is the number of information bits in a 
codeword.
Let $\mathcal{I}\subseteq {\{1, \dots, N\}}$ denote the set 
of indices of the information bits whose compliment set
$\mathcal{I}^c$ denotes the set of 
frozen (non-information bearing) bits. 
Let $\bfu = [u_1, u_2, \cdots, u_N] $ denote the binary 
information vector and let $\bfb = [b_1, b_2, \cdots, b_N]$ denote 
the binary codeword vector. There is a 1-1 mapping 
between $\bfu$ and $\bfb$ defined by $\bfb = \bfu \mathbf{G_N}$ 
in which $\mathbf{G_N}$ is the generator matrix of the polar code. 
Note that $\mathbf{G_N}$ is defined through $\mathbf{G_N} = 
\mathbf{B_N} \mathbf{F}^{\otimes n}$, where $\mathbf{B_N}$ is 
a bit reversal operator defined in \cite{arikan2009channel}, 
and  $\mathbf{F}^{\otimes n}$ denotes $n-$fold Kronecker power of 
polarization kernel 
\begin{equation}
\mathbf{F} = 
\begin{bmatrix}
1 & 0\\
1 & 1 
\end{bmatrix} 
\end{equation}

Using the concept of channel polarization, $N$ identical realization of the 
channel can be transformed  into $N$ parallel 
virtual bit-channels, which become polarized
as either extremely noisy or completely error-free as $N$ tends
to infinity.
Consequently, the crucial step in constructing polar codes is to sort
the virtual bit-channels based on their capacity and to select
the $K$ most reliable ones out of $N$ bit-channels to carry
information bits. The remaining $N - K$ bit-channels 
will carry the frozen bits (set to known values).
As shown in \cite{arikan2009channel}, 
the fraction of reliable virtual channels among all the virtual channels is $K/N$, 
which approaches asymptotically to the channel capacity for 
large $N$. 
As a result, polar codes can achieve channel 
capacity for asymptotically long the codeword length $N\to\infty$ in B-DMCs.

\subsection{Channel Model}
Without loss of generality, we consider an MIMO system with $n$ 
transmit antennas at the BS and $m$
receive antenna at each UE.

For length $N$ polar codeword and $Q$-ary QAM constellation, we can 
transmit one 
codeword via $K_0 = N/Q$ data symbols. For each subcarrier 
index $k$ we 
can define complex QAM symbol $\tilde{x}_{i,k }\in \tilde{Q}$ and
QAM vector $\mathbf{\tilde{x}}[k]=[\tilde{x}_{1,k}\cdots, \tilde{x}_{n,k} ]^T$
as the transmission signal vector.  Let each received 
symbol
be $\tilde{y}_{i,k}\in \mathbb{C}$ and
the received signal vector be $\mathbf{\tilde{y}}[k]=
[\tilde{y}_{1,k},\cdots, \tilde{y}_{m,k} ]^T$. By defining
$\mathbf{\tilde{H}}[k] 
\in \mathbb{C}^{m \times n}$ as
the linear $m\times n$ MIMO channel matrix 
for a flat fading wireless channel of each UE,
we can write the received signal vector as 
\begin{equation}
\label{channel}
\mathbf{\tilde{y}}[k] = \mathbf{\tilde{H}}[k] \mathbf{\tilde{x}}[k] + \mathbf{\tilde{n}}[k],\qquad k=1,\;\cdots,\; K_0
\end{equation}
in which $\mathbf{\tilde{n}}[k]=[\tilde{n}_{1,k}, \cdots, \tilde{n}_{m,k} ]^T \in \mathbb{C}^m$
is the additive white Gaussian noise (AWGN) vector 
whose elements
are i.i.d. complex random variables 
and $\tilde{n}_{i,k} \sim \mathcal{CN} (0, \sigma_n^2)$.
Note that the elements of $\mathbf{\tilde{H}}[k]$ are typically unknown
and are estimated by relying on pilot symbols. Moreover, without loss of generality,
we assume that the channel for each two REs are independent and not the same.

\subsection{Maximum Likelihood Receiver} 
Since the multi-carrier MIMO channels are known or 
estimated at the 
receiver, the optimal maximum likelihood detector (MLD) 
that minimizes the probability of error for each 
transmission $k$ can solve the problem
 \begin{equation}
 \label{MLD}
 \min\limits_{\mathbf{\tilde{x}}\in {\tilde{Q}}^n}
 \sum_{k=1}^{K_0} {\parallel \mathbf{\tilde{y}}[k] - \mathbf{\tilde{H}}[k] \mathbf{\tilde{x}}[k]\parallel^2_2}
 \end{equation}

It should be noted that this
MLD receiver does not yet take 
into consideration the fact that a data symbol vector $\mathbf{\tilde{x}}$ 
must be modulated by bits in an FEC codeword.  In other words, only valid 
FEC codewords should be considered in the MLD receiver. Eliminating
invalid codewords in MLD would have taken into consideration of
the Galois field code constraints 
within the detection stage to minimize the probability of 
producing wrong symbol sequence by the detector to achieve better
performance. Codeword constraints are even more critical when our estimate  of the channel matrix $\mathbf{\tilde{H}}$
is itself inaccurate.

We can denote the bits in each FEC codeword
of length $N$ as $\bfb =[b_1\; b_2\; \cdots b_N]$ which should span $K_0$ data 
vectors, i.e.,
$$ {\cal \tilde{M}}(\bfb)=\{\mathbf{\tilde{x}}[1], \; \mathbf{\tilde{x}}[2],\; \cdots,\; \mathbf{\tilde{x}}[K_0]\}
$$
where ${\cal \tilde{M}}(\cdot)$ denotes the mapping of FEC bits to data symbols
for transmission.  Consequently, the optimum receiver can be written as
\begin{subequations}
\label{FECopt}
\begin{align}
\label{MLD2}
& \min\limits_{\mathbf{b}}
\sum_{k=1}^{K_0}{\parallel \mathbf{\tilde{y}}[k] - \mathbf{\tilde{H}} \mathbf{\tilde{x}}[k]\parallel^2_2}\\
& {\cal \tilde{M}}(\bfb)=\{\mathbf{\tilde{x}}[1], \; \mathbf{\tilde{x}}[2],\; \cdots,\; \mathbf{\tilde{x}}[K_0]\} \label {cm}\\
&\bfb \in {\cal F} \label{cc}
\end{align}
\end{subequations}
where ${\cal F}$ denotes the set of all valid FEC codewords of 
length $N$. 

This exact ML optimization problem is a non-convex optimization problem and 
is highly complex to solve even for moderately long codeword
because it requires exhaustive search over 
all valid set of symbols $ {\cal \tilde{M}}(\bfb)=\{\mathbf{\tilde{x}}[1], \; \mathbf{\tilde{x}}[2],\; \cdots,\; \mathbf{\tilde{x}}[K_0]\}$ that satisfy 
the coding constraints. Such problem is an NP-hard problem 
whose complexity grows exponentially with $n$. 
Furthermore, the constraint that requires 
$\bfb\in {\cal F}$, is defined in Galois field 
and obviously is a non-convex constraint when considering the Euclidean-field
optimization in Eqs.~(\ref{FECopt}).

In the next section we show how to modify the cost 
function and the constraints in (\ref{FECopt})
into an LP optimization problem to derive a unified 
joint receiver.

\section{LP Receiver with Polar Coding Constraints}
\subsection{Redefining the Objective Function}
To formulate a joint LP receiver, our first step is to modify 
the objective function by changing the $l_2$ norm
 in (\ref{MLD}) 
to $l_1$ norm metric. The $l_1$ norm as an optimization
metric has been applied in 
data analysis and parameter estimation because it is robust 
to impulsive noises and other man-made radio
interferences \cite{cui2006linear}. To simplify the $l_1$ notation, we
shall reformulate the problem in real field. 


We assume that our modulation symbols $\tilde{x}_{i,k}$
are from a constellation $\tilde{Q}$ that can be split into
real and imaginary parts such as the most common 
quadrature amplitude constellation (QAM). More specifically,
the real part $\Re\{\tilde{x}_{i,k}\}$ and its imaginary part  $\Im\{\tilde{x}_{i,k}\}$ form the coordinate of a symbol in
the same constellation $Q$. Consequently, 
we can transform our system from complex field to real field by defining
\begin{equation}
\label{real}
\mathbf{y}[k] =
\begin{bmatrix}
\Re{\{\mathbf{\tilde{y}}[k]\}}\\
\Im{\{\mathbf{\tilde{y}}[k]\}}
\end{bmatrix} 
, \mathbf{x}[k] = 
\begin{bmatrix}
\Re{\{\mathbf{\tilde{x}}[k]\}}\\
\Im{\{\mathbf{\tilde{x}}[k]\}}
\end{bmatrix} 
, \mathbf{n}[k] = 
\begin{bmatrix}
\Re{\{\mathbf{\tilde{n}}[k]\}}\\
\Im{\{\mathbf{\tilde{n}}[k]\}}
\end{bmatrix} 
\end{equation}
and
\begin{equation}
\label{realH}
\mathbf{H}[k] = 
\begin{bmatrix}
\Re{\{\mathbf{\tilde{H}}[k]\}} & -\Im{\{\mathbf{\tilde{H}}[k]\}}\\
\Im{\{\mathbf{\tilde{H}}[k]\}}  &  \Re{\{\mathbf{\tilde{H}}[k]\}} 
\end{bmatrix} 
\end{equation}
Given the new notations in the real Euclidean field, we can write our system equation
between the channel input $\mathbf{x}[k]$ and the channel output $\mathbf{y}[k]$ as
\begin{equation}
\label{channel2}
\mathbf{y}[k] = \mathbf{H}[k] \mathbf{x}[k] + \mathbf{n}[k].
\end{equation}

To transform (\ref{cm}) also from complex to real, we define $\cal{M}(\cdot)$ to be the mapping from bit vector $\bfb$ of length $N$ to $
\{\mathbf{x}[1], \mathbf{x}[2], \dots, \mathbf{x}[K_0] \}$ where $\mathbf{x}[k], 1\leq k \leq K_0$ is the set of real transmission symbols defined in (\ref{real}).

We can reformulate our problem into a linear
programming problem with
two sets of generalized vector inequalities by introducing slack variables $e_{i, k}\ge 0,\, 1 \leq i \leq m, 1 \leq k \leq K_0 $. We also define $\mathbf{e}[k]$ to be a vector of these slack variables, i.e. $\mathbf{e}[k] = [e_{k,1}, \cdots, e_{k, m}]$. Consequently we can modify the optimization problem of (\ref{FECopt}) to

\begin{equation}
\begin{aligned}
\label{MILP}
&\text{min} && {\sum_{k=1}^{{K_0}}\sum_{i=1}^{m} e_{i, k} }\\
&\text{  s.t.}  &&\mathbf{H}[k] \mathbf{x}[k] - \mathbf{e}[k]  \preceq  \mathbf{y}[t]\\
& &&-\mathbf{H}[k] \mathbf{x}[k] - \mathbf{e}[k]  \preceq  -\mathbf{y}[k]\\
& && {\cal {M}}(\bfb)=\{\mathbf{x}[1], \mathbf{x}[2],\; \cdots,\; \mathbf{x}[K_0]\}\\
& &&\bfb \in {\cal F} 
\end{aligned}
\end{equation}
Note that $x \preceq y$ denotes element-wise inequality $x_i \leq y_i$.

\subsection{Receiver Integration of Polar Coding Constraints}

Now we describe how to integrate information from the
$\bfb\in {\cal F}$ codeword constraint into the linear programming
optimization of (\ref{MILP}) when polar codes are adopted as FEC. 
Feldman et al. \cite{feldman2005using} proposed one
way to translate linear coding constraints from the Galois field
into certain linear inequality constraints in Euclidean field. 
The basic idea is to relax the constraints imposed by the
parity check matrix of a linear block code 
to define some linear codeword constraints known as 
fundamental polytope $\bar{\mathcal{Q}}$. 
Since the number of transformed linear constraints grow exponentially with the weight of 
the binary parity check matrix, this polytope 
conversion works well for those
low density parity check (LDPC) codes whose
sparsity parity check matrices have very low weights. The integration of such code constraints with detector has been widely investigated in \cite{wang2017galois,wang2018semidefinite} for improved performance. 
Unfortunately, the dense parity check matrix of polar codes 
not only leads to
overwhelming large number of constraints for such polytope $\bar{\mathcal{Q}}$, 
but also exhibits poor decoding performance when applying
linear programming based on such polytope \cite{goela2010lp}. 

\begin{figure}
 \begin{center}
    \includegraphics[scale = 0.36]{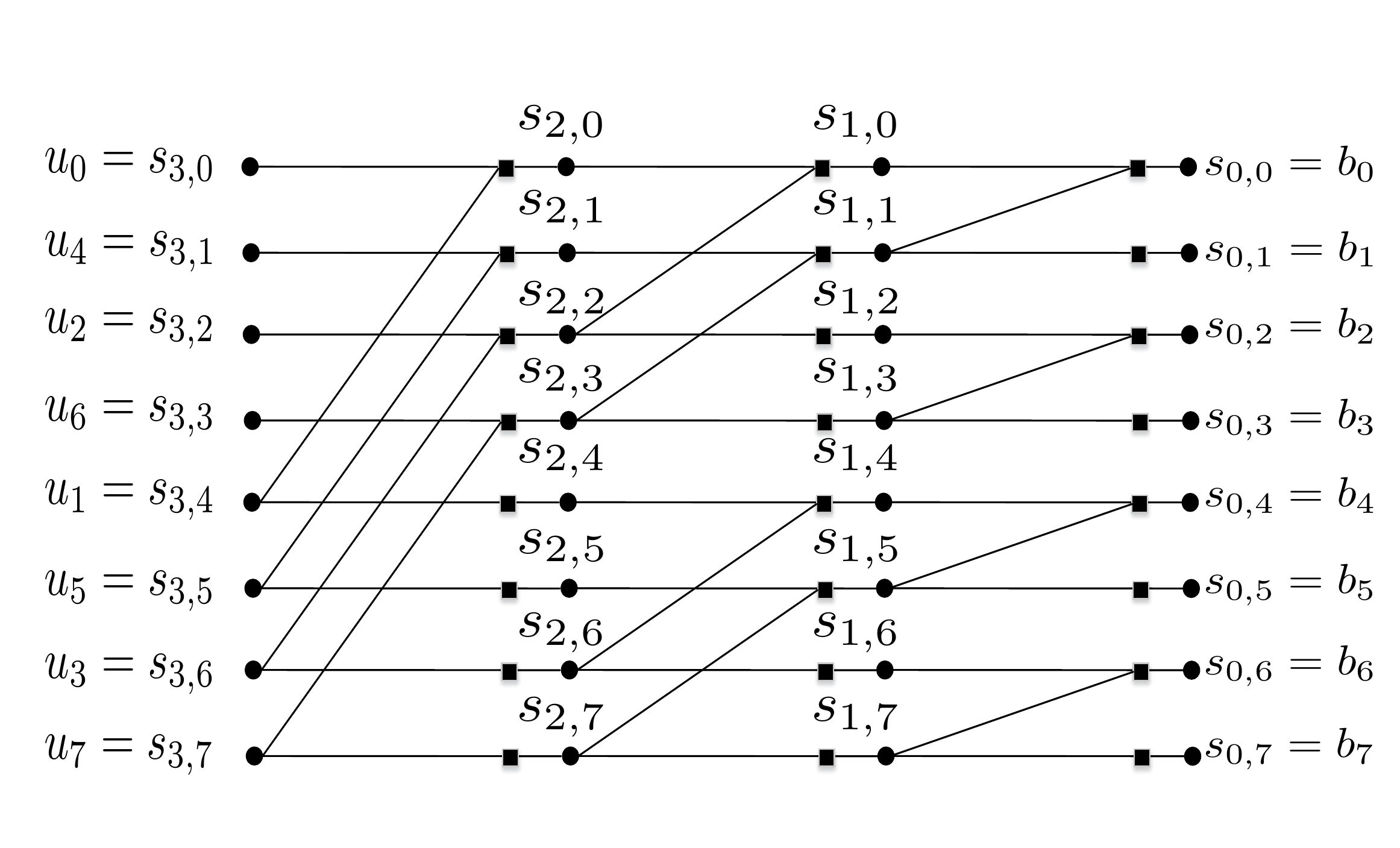}
    \caption{factor graph representation of a polar code with block length $N = 2^3$ }
     \label{fig:fig2}
\end{center}
\end{figure}

Goela et. al \cite{goela2010lp} utilized the recursive 
structure of polar codes that leads to a sparse graph representation with $O(N \log N)$ auxiliary variables, 
where $N$ is the block length. 
Figure \ref{fig:fig2} shows such a factor graph of 
a polar code with block length $N = 2^3$. 
Taking advantage of this factor graph, 
new polytope can be defined in a space of 
dimension $O(N \log N)$ \cite{goela2010lp}.
For this reason, we shall exploit 
this new polytope to
generate a different set of
linear coding constraints that can be incorporated
into the LP receiver of (\ref{MILP}). 

Let us denote corresponding polytope as $\mathcal{P}$. 
The graph of Figure~\ref{fig:fig2} shows how a polar codeword $\bfb$ can be constructed from binary vector $\mathbf{u}$ by a 1-1 mapping through 
the generator matrix $\mathbf{G_N}$,  $\bfb = \mathbf{u} \mathbf{G_N}$. 
The circle nodes on the graph represent a total 
of $N (1 + \log N)$ binary variables and 
the square nodes represent the check nodes. 
If all the check nodes are satisfied,
then $\bfb$ is a valid 
codeword. 

An example of a check node constraint in Figure \ref{fig:fig2} is  
\[ u_0 \oplus u_1 \oplus s_{2,0} = 0\]
where $\oplus$ denotes modulo-2 addition. 
To define the relaxed polytope $\mathcal{P}$, we let the variables in the graph be real variables instead of binary. Note that each constraint involves only
either 3 or 2 variables. Therefore, for each check 
node $j \in \mathcal{J}$ 
with 3 neighbors $\mathcal{N}(j) = \{a_1, a_2, a_3\}$, the local minimal 
convex polytope of check node $j$ is $\mathcal{P}_j$, 
which can be 
very simply defined by below linear inequalities
\begin{equation}
\begin{aligned}
&0 \leq a_1\leq a_2 + a_3,\\
&0 \leq a_2\leq a_3 + a_1,\\
&0 \leq a_3\leq a_1 + a_2,\\
& a_1+a_2 + a_3 \leq 2\\
\end{aligned}
\end{equation}
For each check node $j \in \mathcal{J}$ with only two neighbors $\mathcal{N}(j) = \{a_1, a_2\}$, the local polytope $\mathcal{P}_j$ is defined by
\begin{equation}
\begin{aligned}
&a_1 = a_2\\
&0 \leq a_1 \leq 1\\
&0 \leq a_2 \leq 1\\
\end{aligned}
\end{equation}
Moreover, we denote the cutting plane $\cal{T}$ 
as defined by setting all frozen variables with
indices belonging to $\mathcal{I}^c$ to zero. 
In summary, the polytope $\mathcal{P}$ is the 
intersection of all local polytopes plus
the cutting plane $\mathcal{T}$ via
\begin{equation}
\label{relax}
\mathcal{P} = \left(\bigcap_j \mathcal{P}_j \right) \cap \mathcal{T}
\end{equation}

Therefore, we can write down the linear coding 
constraints by enforcing all the variables of the 
factor graph to comply with the polytope $\mathcal{P}$, 
i.e. $\mathbf{s} \in \mathcal{P}$ where $\mathbf{s}$ 
denotes all the variables of the factor graph. These
constraints can be incorporated into (\ref{MILP}) 
as relaxed version of (\ref{cc}).
Therefore, the final formulation for the LP 
receiver can be simply 
written as
\begin{subequations}
	\label{LP3}
\begin{eqnarray}
&\hspace*{-8mm}\text{min} & {\sum_{k=1}^{{K_0}}\sum_{i=1}^{m} e_{i, k} }\\
&\hspace*{-8mm}\text{  s.t.}  &
\quad \mathbf{H}[k] \mathbf{x}[k] - \mathbf{e}[k]  \preceq  \mathbf{y}[k], \quad k=1,\cdots, K_0\\
& &-\mathbf{H}[k] \mathbf{x}[k] - \mathbf{e}[k]  \preceq  -\mathbf{y}[k], \quad k=1,\cdots, K_0\\
& & {\cal {M}}(\bfb)=\{\mathbf{x}[1], \mathbf{x}[2],\; \cdots,\; \mathbf{x}[K_0]\}\\
& & \mathbf{s} \in \mathcal{P} \subseteq [0, 1] ^ {N (1 + \log N)}
\end{eqnarray}
\end{subequations}

Note that all $N (1 + \log N)$ variables in $\mathbf{s}$ are optimization variables in the problem of (\ref{LP3}). 
As a result, the detector generates an estimate for each of these variables including $\bfb$. 
We denote the estimated bit vector as 
$\hat{\bfb} = [\hat{b}_1\; \hat{b}_2\; \cdots \hat{b}_N]$.
Each element $0 \leq \hat{b}_i \leq 1$ can be used to derive
the likelihood of $b_i$ to be 0 or 1. Once the
log-likelihood ratio is generated by 
our joint LP detector, a soft-input decoder 
such as SC or SCL can be used to
further decode $\hat{\bfb}$ to produce a final output $\hat{\mathbf{u}}$
which are the receiver outputs of the uncoded source bits.
This process describes the joint detection and 
decoding algorithm executed by our proposed LP receiver.

\section{Blind Detection Scheme}
\subsection{PDCCH Candidates and Fractional Metric}
In the system diagram shown in Figure \ref{fig:fig1}, 
$C$ PDCCH candidates are possibly received 
at the joint LP detector at the same time. 
The goal for the detector is first to estimate 
the bit sequence from the detected
complex symbols reliably. 
For each decoded codeword which is a candidate
PDCCH for the UE, the UE shall check
its CRC after unmasking by its RNTI. 
Based on the CRC checking result, the 
UE determines whether or not the decoder output is 
a relevant DCI for the UE. 
The more candidates the detector pass to the
decoder for decoding and CRC-checking, the
higher the receiver complexity.

To reduce the receiver complexity, latency,
and power consumption, 
it is important for the receiver to
eliminate as many false candidates as 
possible before passing them to the 
decoding stage. 
However, it is important to note
 that the penalty of eliminating a correct 
candidate is much more severe than the benefit of 
rejecting a false candidate, since by missing
the DCI, a UE would lose its ability to obtain
its data payload in PDSCH that is destined to
the UE.  Such loss of DCI due to the failure of
finding the right PDCCH will have a severe impact 
on the overall system throughput.
Thus, we must take this effect into 
account when designing our proposed 
low complexity receiver 
algorithm that reject the false PDCCH candidates.

Note that because of RNTI unmasking, DCI for
other UEs would appear to be a random bit block to 
the receiving UE because the unmasking RNTI would be different from the RNTI that is used to mask the
DCI by the gNB transmitter. 
In order to help UE distinguish good PDCCH candidates
from unlikely candidates,  we propose a new metric
that takes smaller values if the received block is a 
valid polar code block against the case when the 
received bit sequence block appears to be random bits. 

Let $\mathbf{s}^\star = [\mathbf{s}^\star_1, \cdots, \mathbf{s}^\star_{N(1+\log N)}]$, $0 \leq \mathbf{s}^\star_i \leq 1 $ denote the solution sequence of the
LP detector for the whole factor graph as defined by (\ref{LP3}). We define a metric $f$ 
\begin{equation}
\begin{aligned}
f = {\parallel \mathbf{s}^\star - [\mathbf{s}^\star] \parallel }_1
\end{aligned}
\end{equation}

The feature metric $f$ 
indicates the $l_1$ distance between the LP solution $\mathbf{s}^\star$ and the closest vertex in the
$N (1 + \log N)$-hypercube $[0,1]^{N(1+\log N)}$. 
Therefore, conceptually, $f$ is a measure 
of how \emph{fractional} the LP solution is, denoting 
the dominance of fractional solution 
as discussed in \cite{feldman2005using}. 
We know that $\mathbf{s^\star} \in \mathcal{P} 
\subseteq [0, 1] ^ {N (1 + \log N)}$; however, the 
transmitted bit sequence in downlink corresponds 
to a vertex of the hypercube. Therefore, without relaxing
the bit variables to $[0,1]$, $\mathbf{s^\star}$ must
be on a vertex and, consequently, $f$ would 
have been zero. 
The closer the LP solution is to a vertex of
the hypercube, the more reliable the solution is. 

If the bit sequence transmitted through the channel is 
a valid polar codeword, and also if the SNR is high and
channel state information is perfectly known, then 
the detector will be able to retrieve the true
bit sequence from the BS transmitter. The bit 
sequence that was actually sent can both minimize 
the objective function of (\ref{MLD2}) and also 
satisfy the polar code constraints simultaneously, i.e. $\mathbf{s^\star} \in \mathcal{P}$. 
Since $\mathbf{s^\star}$ in this case would only
consist of binaries integers, 
$f$ would be zero as well.
For larger channel noise increases, channel estimation
would suffer. Thus, the LP solution would differ
from the actual bits sent and therefore, the LP solution
is likely to be dislodged from the vertex of the 
hypercube toward a fractional solution. 
However, under moderate to high SNR, 
the LP solution is still expected to be 
near the transmitted bit sequence. Hence,
its distance from a hypercube vertex and 
consequently, its fractional metric $f$
should be small and near zero.

On the other hand, consider the scenario that
the DCI for a different UE is being examined as
a PDCCH candidate.  With the RNTI mismatch between
masking and unmasking, a random bit sequence is 
being generated as the output
by the detector as opposed to a valid polar codeword.  
In this case, although the correctly detected
bit sequence can actually minimize the objective 
function of the MIMO detection at high SNR, 
this sequence cannot satisfy the polar code constraints 
of LP detector for this particular UE whose RNTI differs
from the masking RNTI used by the transmitter.
Hence, the transmitted sequence
does not belong to the polytope $\mathcal{P}$. 
The LP detector has to find a solution that minimizes 
the objective function while remaining in the
feasible set of (\ref{LP3}). 
Therefore, the LP detector solution $\mathbf{s^\star}$ 
would not be necessarily close to one of the vertexes of the hypercube and can lie anywhere within 
the feasible set. Therefore, the metric 
$f$ in this case
is unlikely to be small.

To demonstrate the comparative values 
of $f$ for the two cases 
discussed above, 
we present the empirical probability density functions
of $f$ under different channel conditions
quantified by different levels of SNR. 
Let $\mathcal{H}_1$ denote the event that a valid polar 
code was
transmitted sent and received by the UE, and let $\mathcal{H}_0$
denote the event that a random bit block was received by
the UE. 
Furthermore, we consider both the case of perfect channel estimate
and imperfect channel estimate at the receiver. 
Specifically, to model imperfect channel estimates,
our estimate of the channel matrix $\mathbf{\hat{H}}$ is assumed to be

\begin{equation} 
\label{err}
 \mathbf{\hat{H}} = \mathbf{H} + \mathbf{E} 
 \end{equation}
 
where $\mathbf{H} = [h_{ij}] \in \mathbb{R}^{2m \times 2n}$ is the real transformation of the complex channel matrix $\mathbf{\tilde{H}}$, based on (\ref{realH}). Therefore, $\mathbf{H} = [h_{ij}] \in \mathbb{R}^{2m \times 2n}$ consists of i.i.d. elements such that $h_{i,j} \sim \mathcal{N} (0, 1)$. The estimation error matrix $\mathbf{E} = [e_{ij}] \in \mathbb{R}^{2m \times 2n}$ 
also consists of i.i.d. random elements such that 
$e_{i,j} \sim \mathcal{N} (0, \alpha\frac{\sigma_n^2}{2})$, where $\sigma_n^2$ is the noise variance and $\alpha \geq 1$ depends on whether there are enough pilot symbols to estimate the channel accurately or not. In our simulations we investigate two cases of $\alpha = 1$ (enough pilot symbols) and $\alpha = 2$ (short pilot length leading to estimation error variance to be double the noise variance).

\begin{figure*}[!t]
\subfloat[SNR = 2dB, accurate channel]{\includegraphics[scale=0.41]{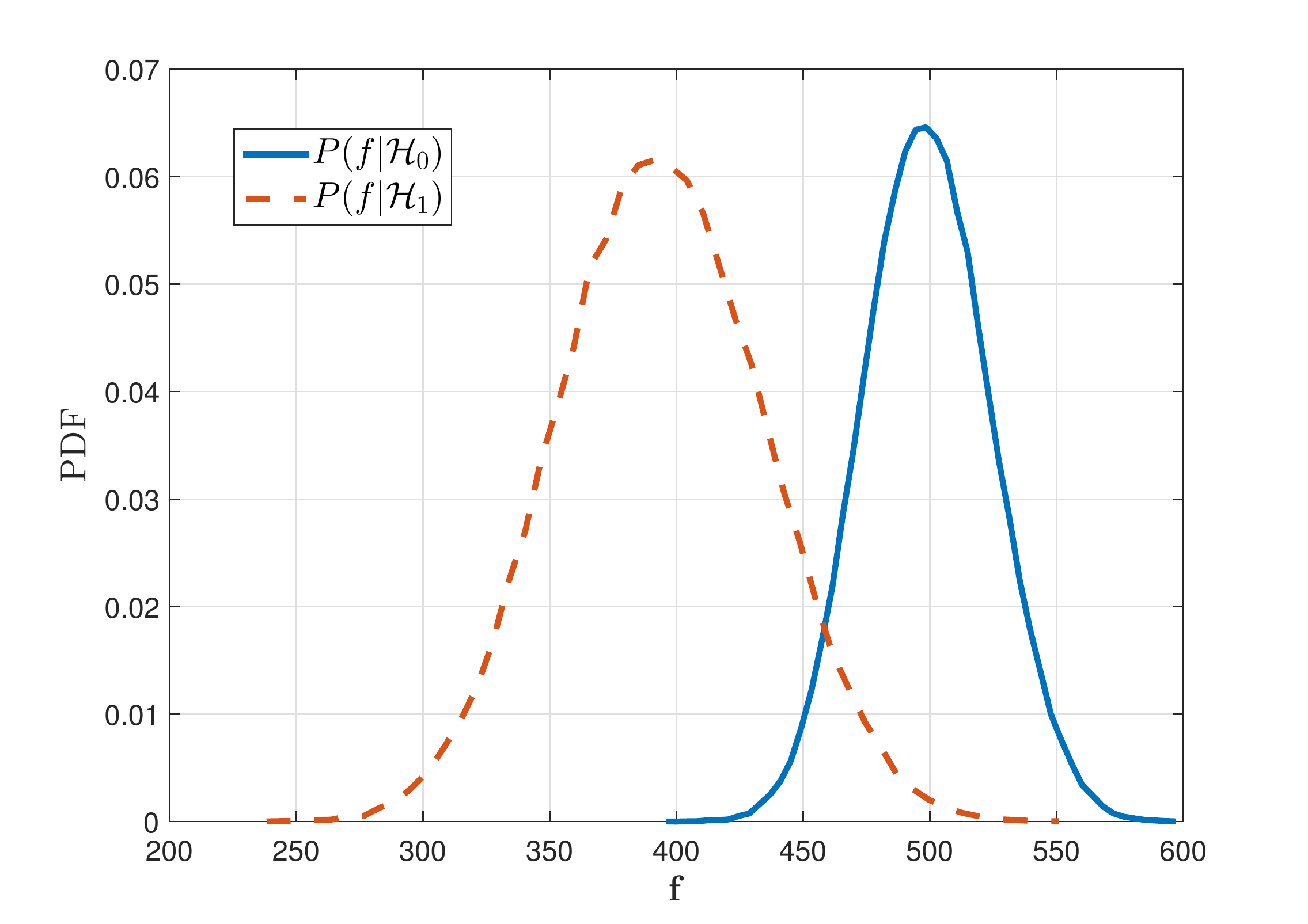}
\label{empirical_accurate_2db}}
\subfloat[SNR = 5dB, accurate channel]{\includegraphics[scale=0.41]{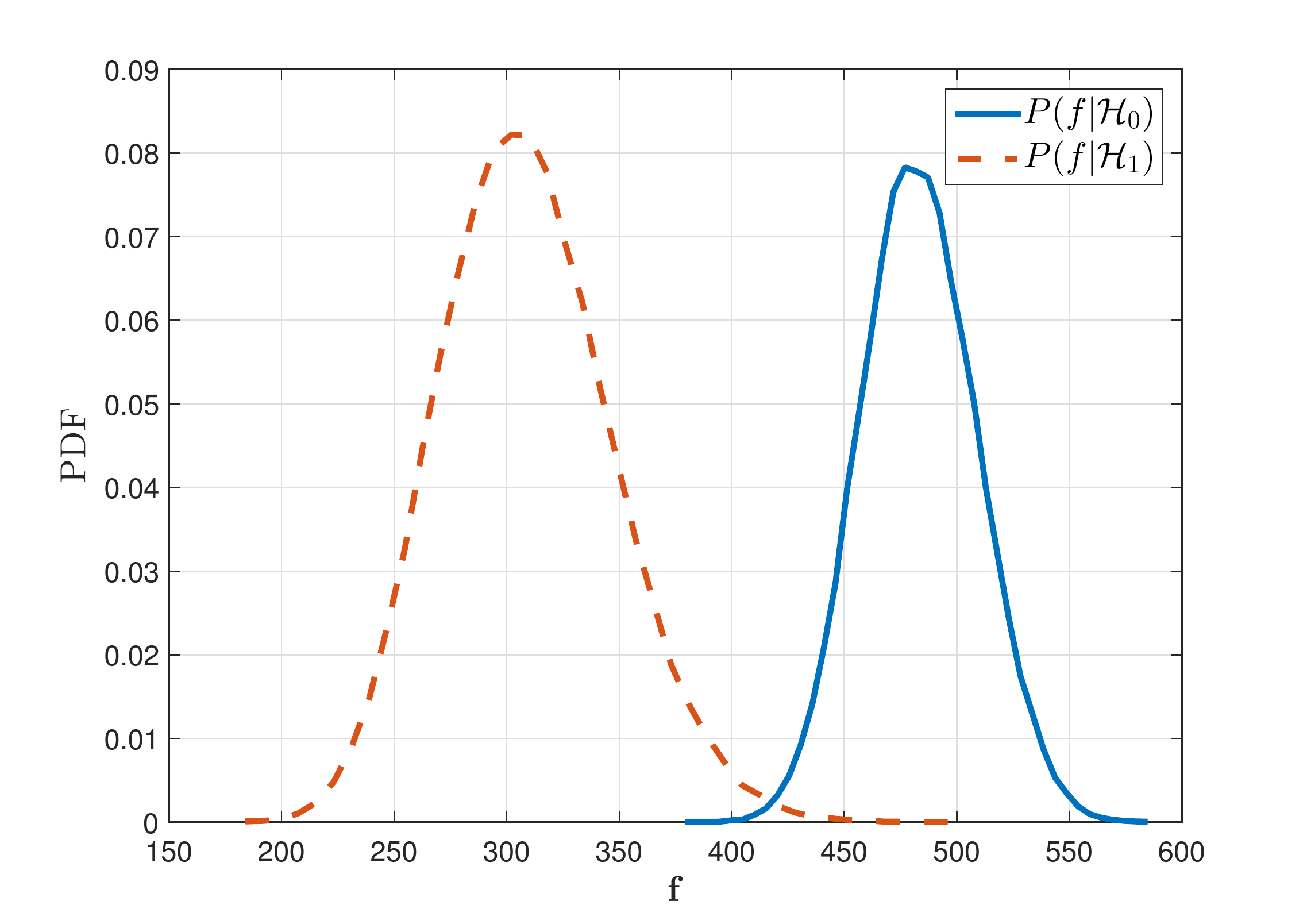}
\label{empirical_accurate_5db}}

\subfloat[SNR = 8dB, accurate channel]{\includegraphics[scale=0.41]{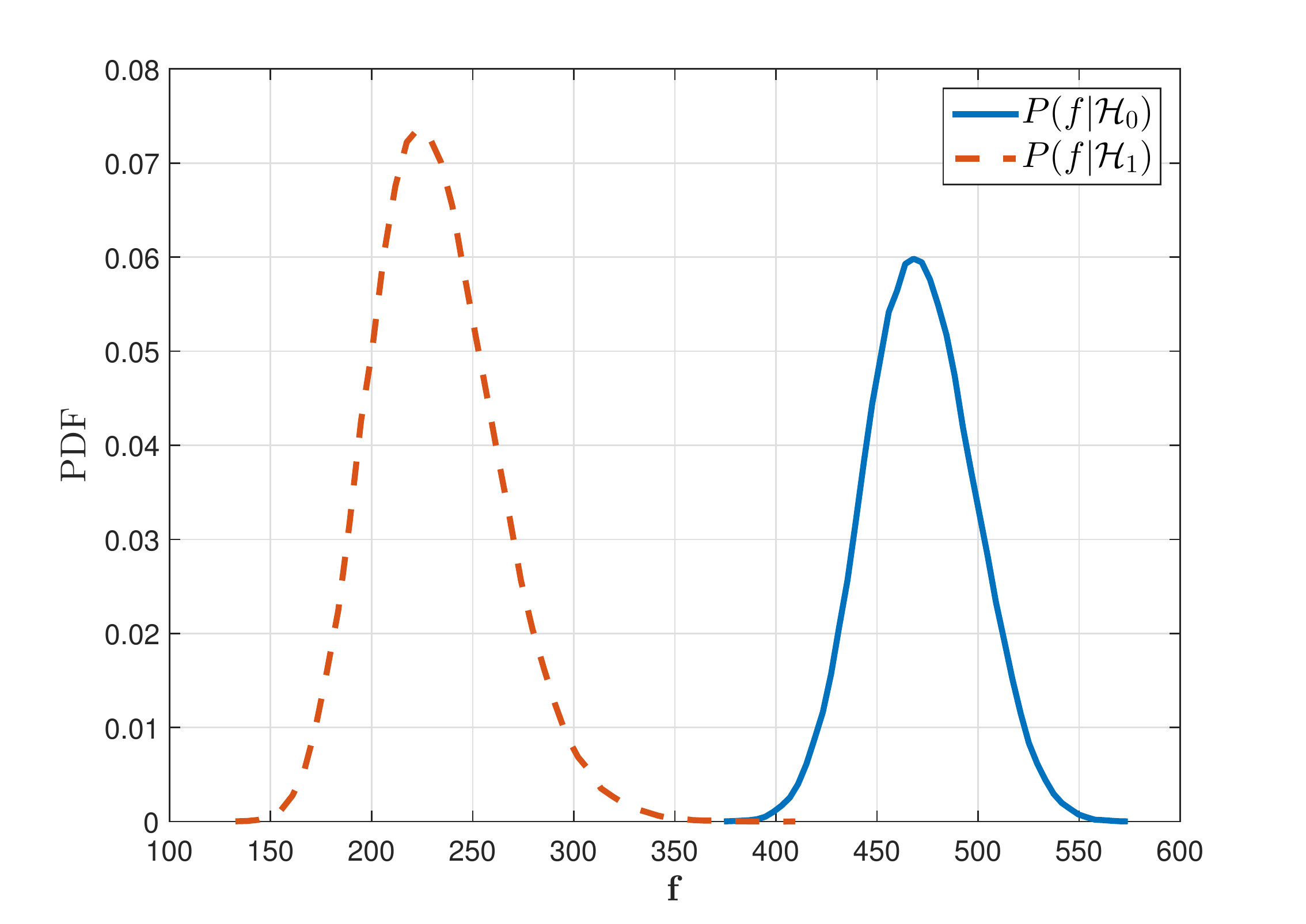}
\label{empirical_accurate_8db}}
\subfloat[SNR = 5dB, inaccurate channel]{\includegraphics[scale=0.41]{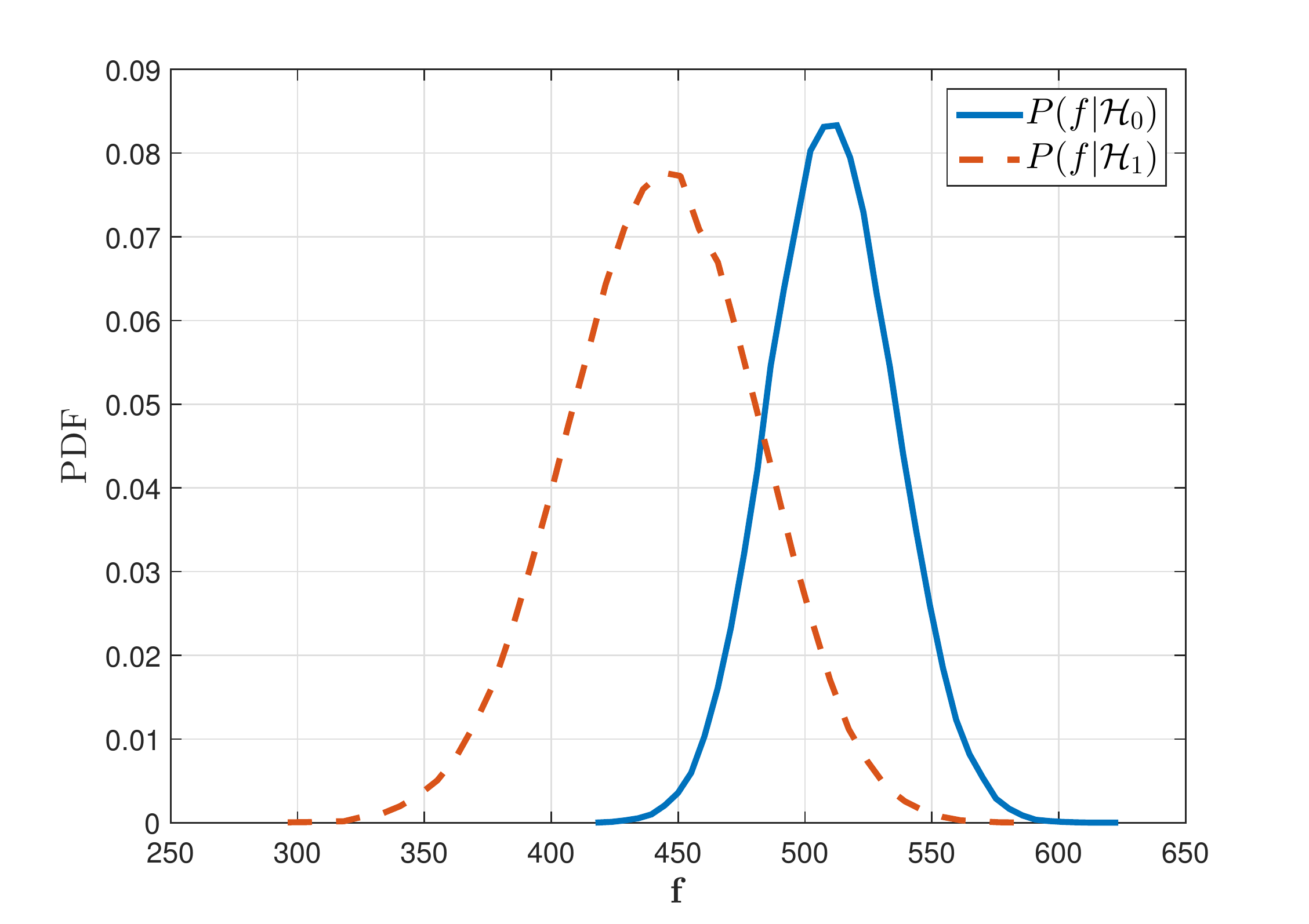}
\label{empirical_inaccurate_5db}}

\subfloat[SNR = 8dB, inaccurate channel]{\includegraphics[scale=0.41]{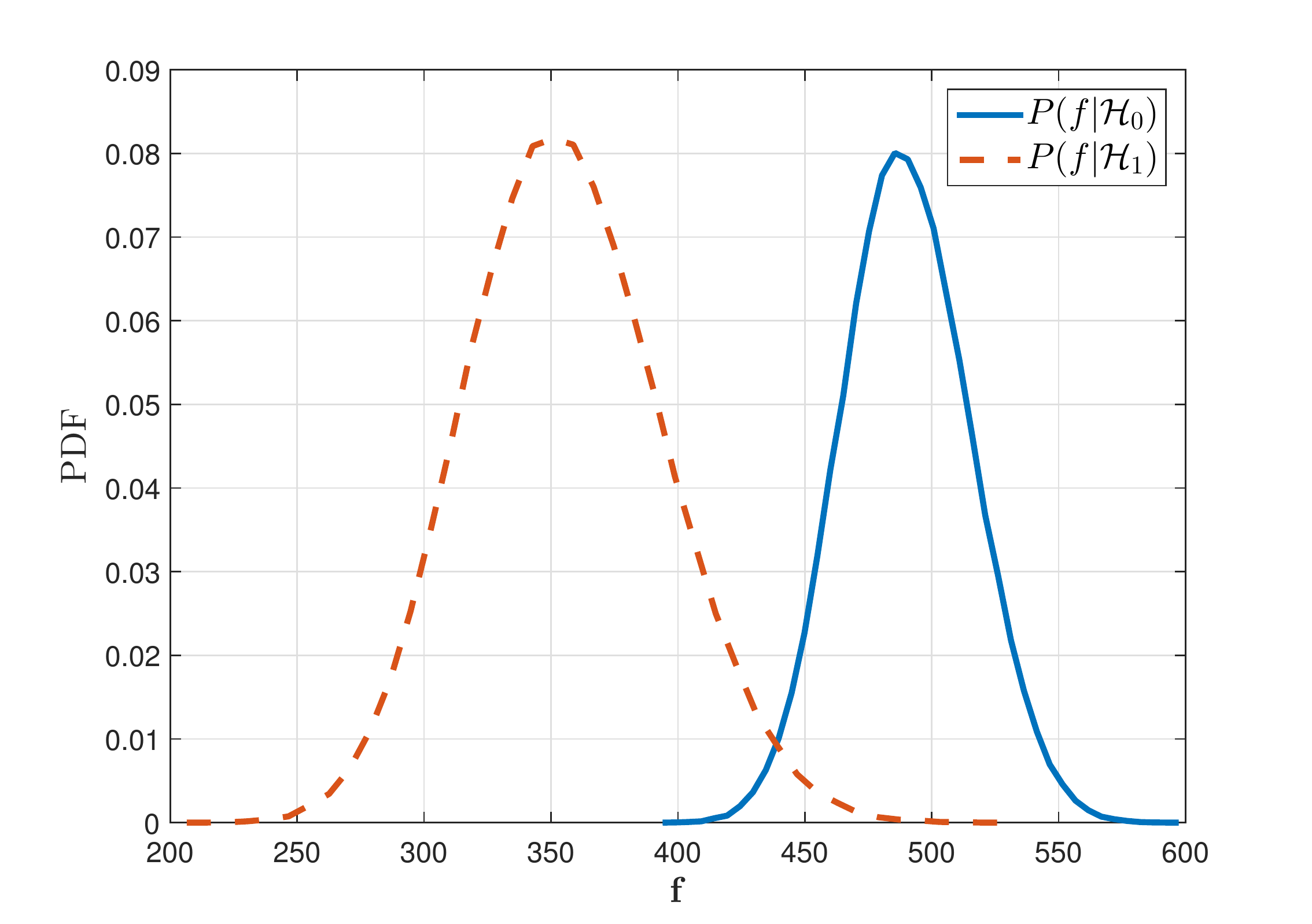}
\label{empirical_inaccurate_8db}}
\subfloat[SNR = 11dB, inaccurate channel]{\includegraphics[scale=0.41]{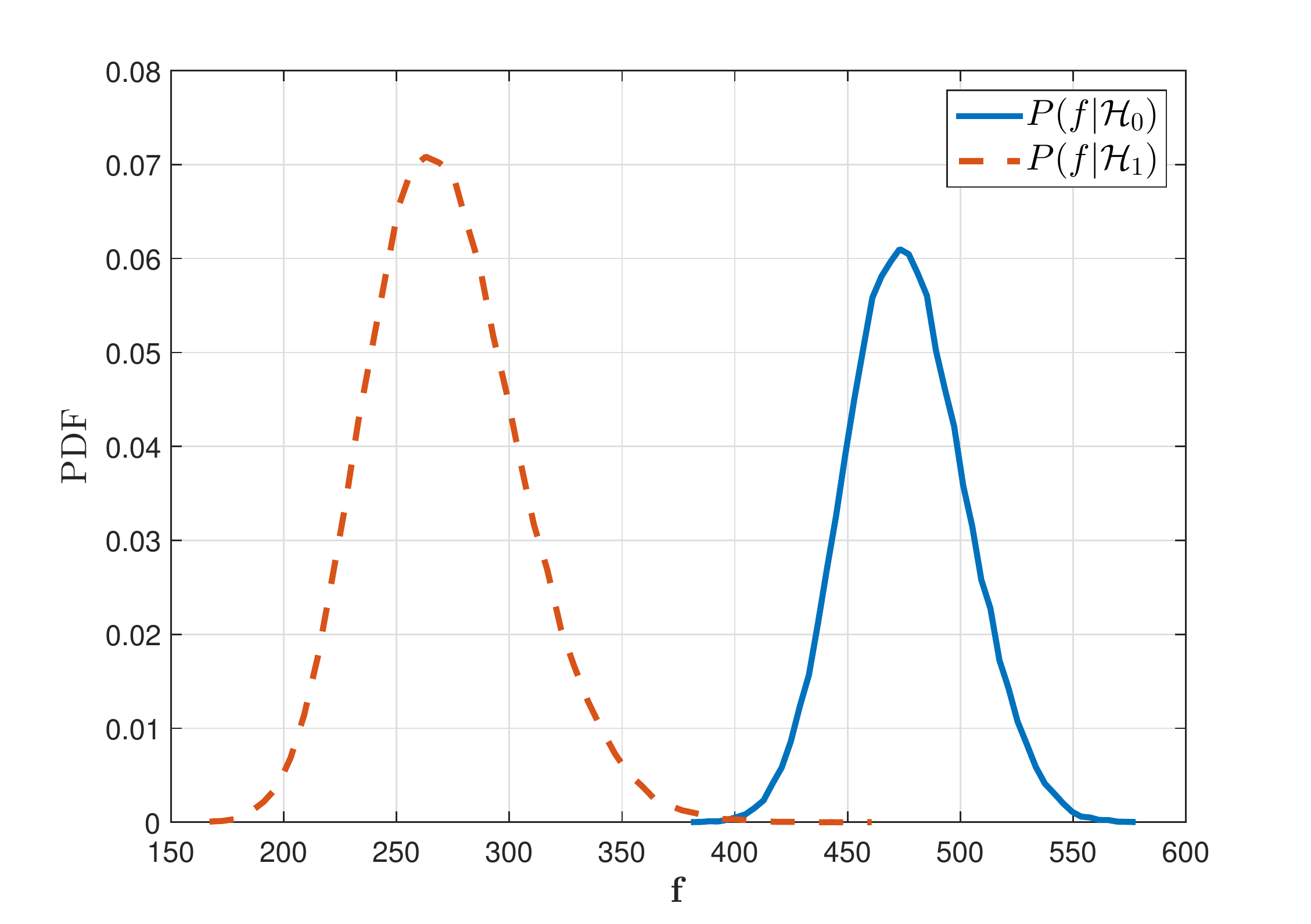}
\label{empirical_inaccurate_11db}}
\vspace{5mm}
\caption{probability density function of $f$ under $\mathcal{H}_0$ and $\mathcal{H}_1$
 for different SNRs when channel matrix is accurate/inaccurate}
\label{fig_sim}
\end{figure*}

As shown in 
Figures \ref{fig_sim}, the conditional
distributions $P(f | \mathcal{H}_0 )$ and $P(f | \mathcal{H}_1 )$ are fully distinguishable from each other.
In particular, $f$ under $\mathcal{H}_1$ tends to be
generally smaller whereas $f$ under $\mathcal{H}_0$
tends to be much larger. As expected, 
the separation between the two conditional probability density
functions become larger for higher SNRs. 
Moreover, the separation becomes less pronounced
under inaccurate channel estimation.
From these numerical tests, it is clear that the metric
$f$ is an effective measure on whether
a detection output sequence is a likely candidate polar codeword.

Qualitatively, 
$P(f | \mathcal{H}_0 )$ is hardly affected by the
changing SNR, since the solution under $\mathcal{H}_0$ 
can be anywhere in the feasible polytope $\mathcal{P}$.
Thus, for event $\mathcal{H}_0$, the SNR would not change 
how far the solution would lie from a vertex of the hypercube.
On the other hand, under $ \mathcal{H}_1$, the
detector at higher SNR becomes more 
successful in correctly estimating the true codeword,
thereby leading to smaller values of  $f$. 
For similar reason, higher 
SNR is needed in case of imperfect channel for the metric $f$ to become smaller. 

Figure \ref{fig_sim} is a clear indication that 
the \emph{fractional metric}
$f$ can be used to differentiate between the 
two cases at moderate to high SNR.  We shall take advantage of $f$ to eliminate false candidates and consequently reduce significantly the number of candidates 
before invoking SC/SCL decoder. 

\subsection{Effective PDCCH Candidate Trimming}

Due to the large number of PDCCH candidates generated
by the UE detector, we need an effective algorithm to eliminate 
those least likely candidates in order to lower the
receiver complexity and power consumption. 
For each PDCCH candidate generated by the joint LP receiver, 
we can efficiently compute its corresponding 
fractional metric $f$.
Considering the fractional metric, we propose to
view the problem of candidate selection
as a clustering problem in which we attempt to decide 
whether all $C$ candidates are derived from probability 
distribution function $P(f | \mathcal{H}_0 )$ or 
there exists one candidate in $C$ is 
derived from $P(f | \mathcal{H}_1)$. 

If we can determine with high probability that one 
candidate is likely drawn from $P(f | \mathcal{H}_1)$,
then we can pass the candidate codeword to the decoder 
and eliminate all other $C-1$ candidates. 
If we cannot have high confidence to select the most
likely candidate, then we should pass a subset of
$C_1$ likely candidates from $C$ candidates
to the decoder and let CRC indicate which one is corresponding to the UE. 

Let us denote $\mathbf{a} =  [f_1, \cdots, f_C]$.
as the array that contain the fractional metrics 
for the $C$ candidates. We propose the following 
algorithm to decide which PDCCH candidates should be 
passed to the decoder.

\vspace{2mm}
\hrule
\vspace{1mm}
\begin{algorithmic}
\label{alg}
\STATE $f_{\min} \gets \text{minimum of $\mathbf{a}$}$
\STATE $\mathbf{\tilde{a}}\gets \text{eliminate $f_{\min}$ from $\mathbf{a}$}$
\STATE $\sigma \gets \text{standard deviation of $\tilde{\mathbf{a}}$}$
\STATE $\mu\gets \text{mean of $\tilde{\mathbf{a}}$}$
\IF {$f_{\min}\leq \mu - \beta \times \sigma$} 
        \STATE $\text{send the candidate corresponding to $f_{\min}$ to the decoder}$
        \STATE $\text{eliminate all other candidates}$
        
\ELSE
        \STATE$ \text{send all $C$ candidates to the decoder}$
\ENDIF 
\end{algorithmic}
\vspace{1mm}
\hrule
\vspace{2mm}

The algorithm would generate a wrong candidate 
only if, first of all, $P(f | \mathcal{H}_0)$ produces an $f$ metric that deviates from its mean by $\beta \times \sigma$, where $\sigma$ is the empirical standard deviation of $P(f | \mathcal{H}_0)$. Moreover, this $f$ value 
must be smaller than what is drawn from $P(f |\mathcal{H}_1)$. 
As it can be seen from Figure (\ref{fig_sim}),
the probability of the latter decreases with increasing SNR 
and decreasing channel estimation error. The 
probability of the first event is also increasingly
unlikely as we select increasingly larger $\beta$ in our algorithm.

Our proposed algorithm has a design parameter $\beta$. 
The higher the value of $\beta$, the more conservative we
are for declaring one valid polar codeword among the candidates. 
Therefore, probability of missed detection should be
lower by selecting a larger $\beta$. 
On the other hand, a larger $\beta$ also means a higher
likelihood that detector passes all $C$ candidates to the decoder,
thereby leading to higher cost in terms of
computation complexity, latency, and energy consumption.  
Therefore, by varying $\beta$ we can trade the probability 
of missed detection on one hand for receiver cost reduction
on the other hand.

Based on the empirical tests, we observe that the
conditional PDF $P(f |\mathcal{H}_i)$ are approximately
Gaussian. Thus, we can find an upper bound for the missed detection probability

\begin{equation}
\begin{aligned}
P(\text{missed detection})    &= P ((f_0 \leq f_1) \cap (f_0 \leq \mu - \beta \sigma)) \\
                                         &\leq P(f_0 \leq \mu - \beta \sigma)\\ 
                                         &= 1 - Q(-\frac{\mu - \beta\sigma - \mu}{\sigma}) = Q(\beta)
\end{aligned}
\end{equation}

Where $f_1$ is assumed to be the fractional metric of the right candidate and $f_0$ is the smallest fractional metric among all the false candidates, 
that are drawn from $P(f |\mathcal{H}_1)$ and $P(f |\mathcal{H}_0)$ respectively.
Owing to the importance of receiving the correct control
information DCI
in order to be able to decode DL-SCH, we target a 
probability of missed detection below $10 ^ {-6} $.
Since $Q(5) = 2.87 \times 10^{-7}$, $\beta \geq 5$ would be a reasonable choice.
We will show that this requirement will meet our expectations 
in our simulation results.

\section{Simulation results}

We first present a set of simulation tests and results to demonstrate the
performance of the proposed joint LP receiver in terms of bit error 
rate (BER) and block error rate (BLER). In particular, we will compare the joint LP receiver with a conventional decoupled MMSE detector in which detection and decoding are performed sequentially. Moreover, we will test our proposed algorithm in \ref{alg} to demonstrate its capability to correctly identify the right PDCCH candidate for the UE.
Throughout this section, we utilized the MOSEK solver \cite{andersenmosek} 
to solve the LP in our simulations. 

\subsection{BER and BLER}
In our simulation tests, we adopt both
a $4 \times 4$ spatial multiplexing MIMO model and a $4 \times 1$ transmit diversity MIMO wireless
model with QPSK modulated PDCCH over a flat Rayleigh fading channel. 
For PDCCH, we adopt a polar code of rate = 0.5 with length $N=128$. 
At the receiver, the flat fading MIMO channel is estimated for each subcarrier 
that also includes channel estimation errors as 
characterized in (\ref{err}).
We calculate the BER and BLER for both the decoder output 
and the output of the detector by hard slicing the soft information of its output.

\begin{figure}
 \begin{center}
    \includegraphics[scale = 0.4]{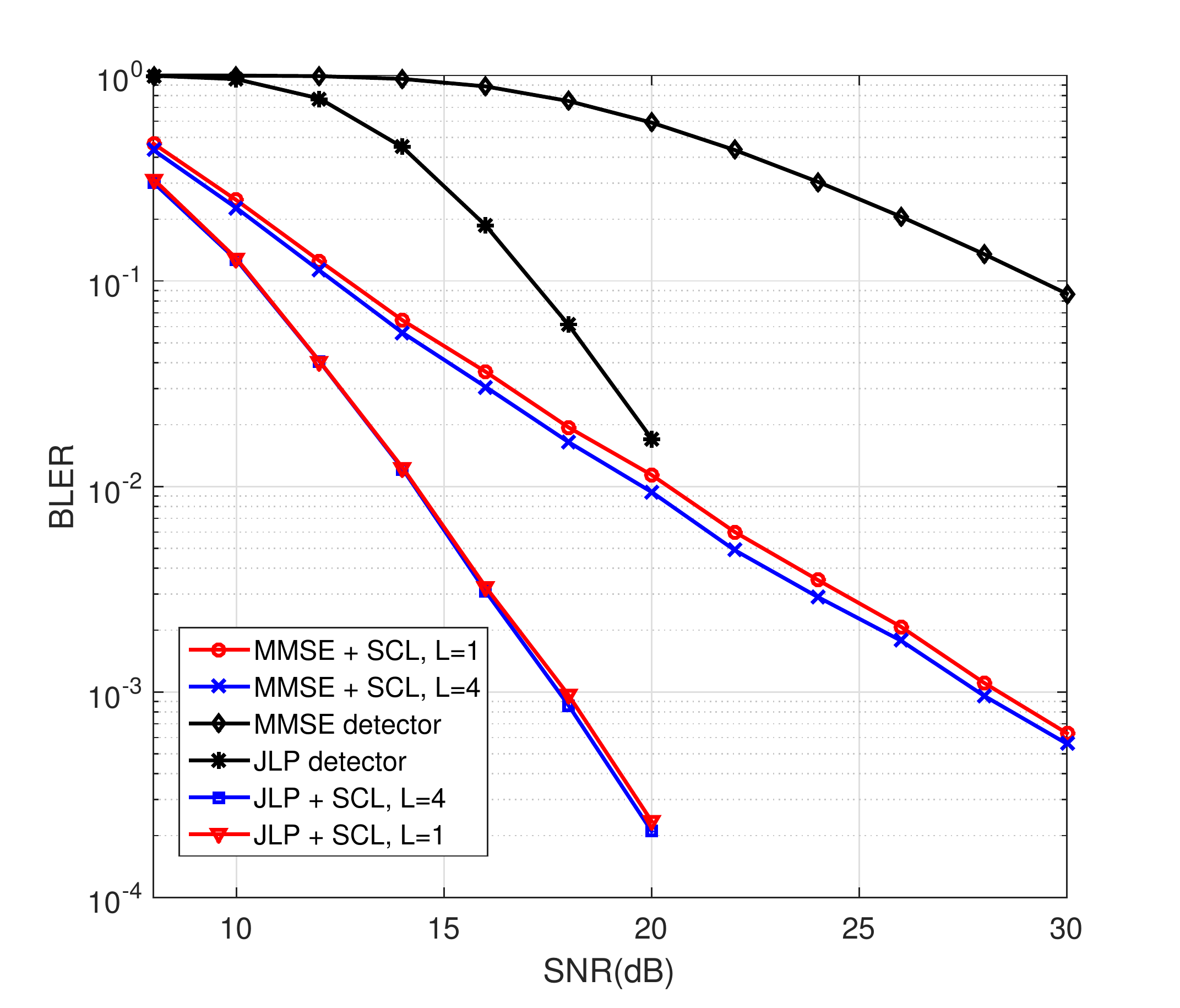}
    \includegraphics[scale = 0.4]{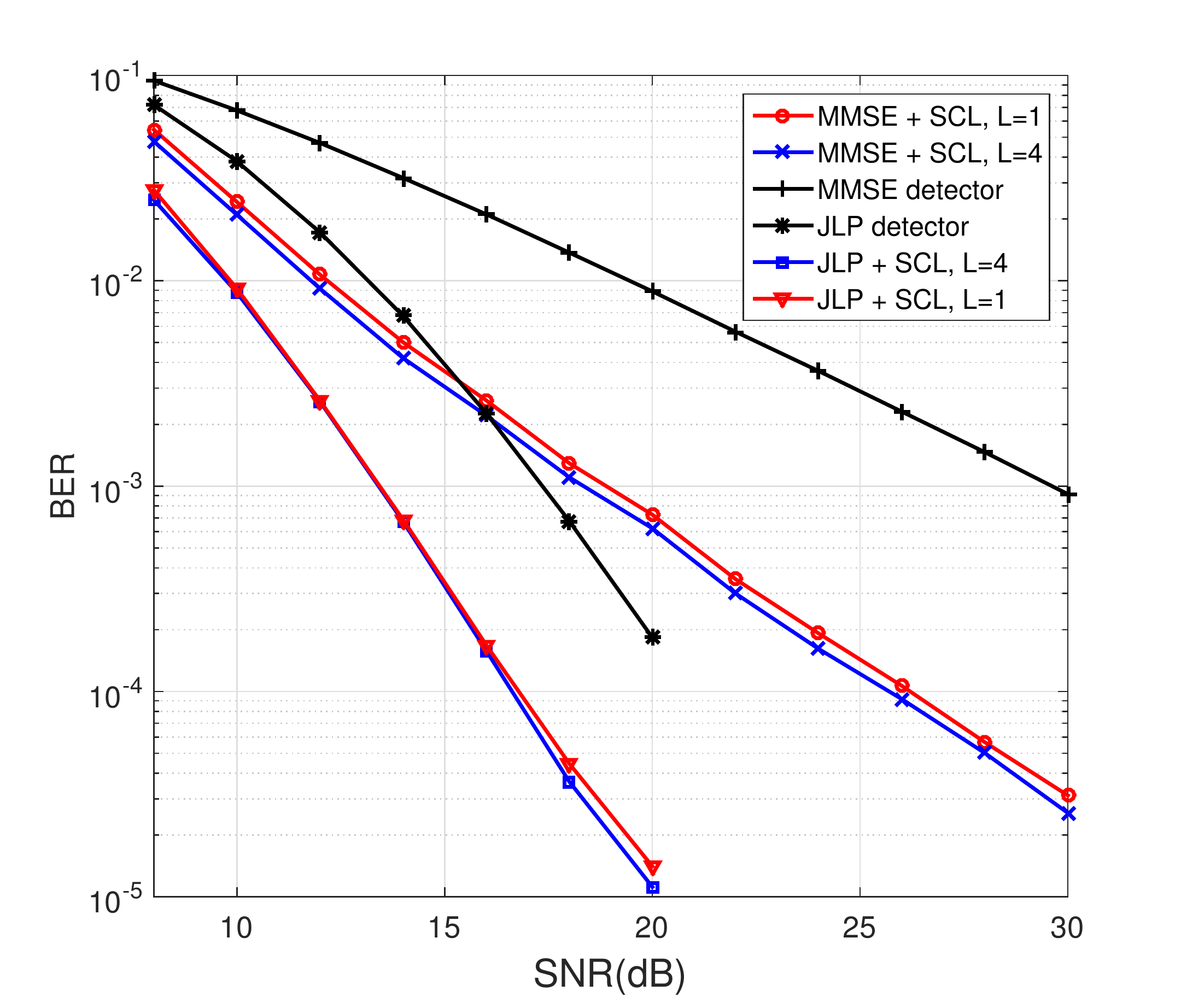}
    \caption{Performance comparison of joint LP detector using code constraints versus decoupled MMSE, $\alpha = 1$, $4 \times 4$ MIMO}
     \label{fig:fig3}
\end{center}
\vspace*{-2mm}
\end{figure} 
\begin{figure}
 \begin{center}
     \includegraphics[scale = 0.4]{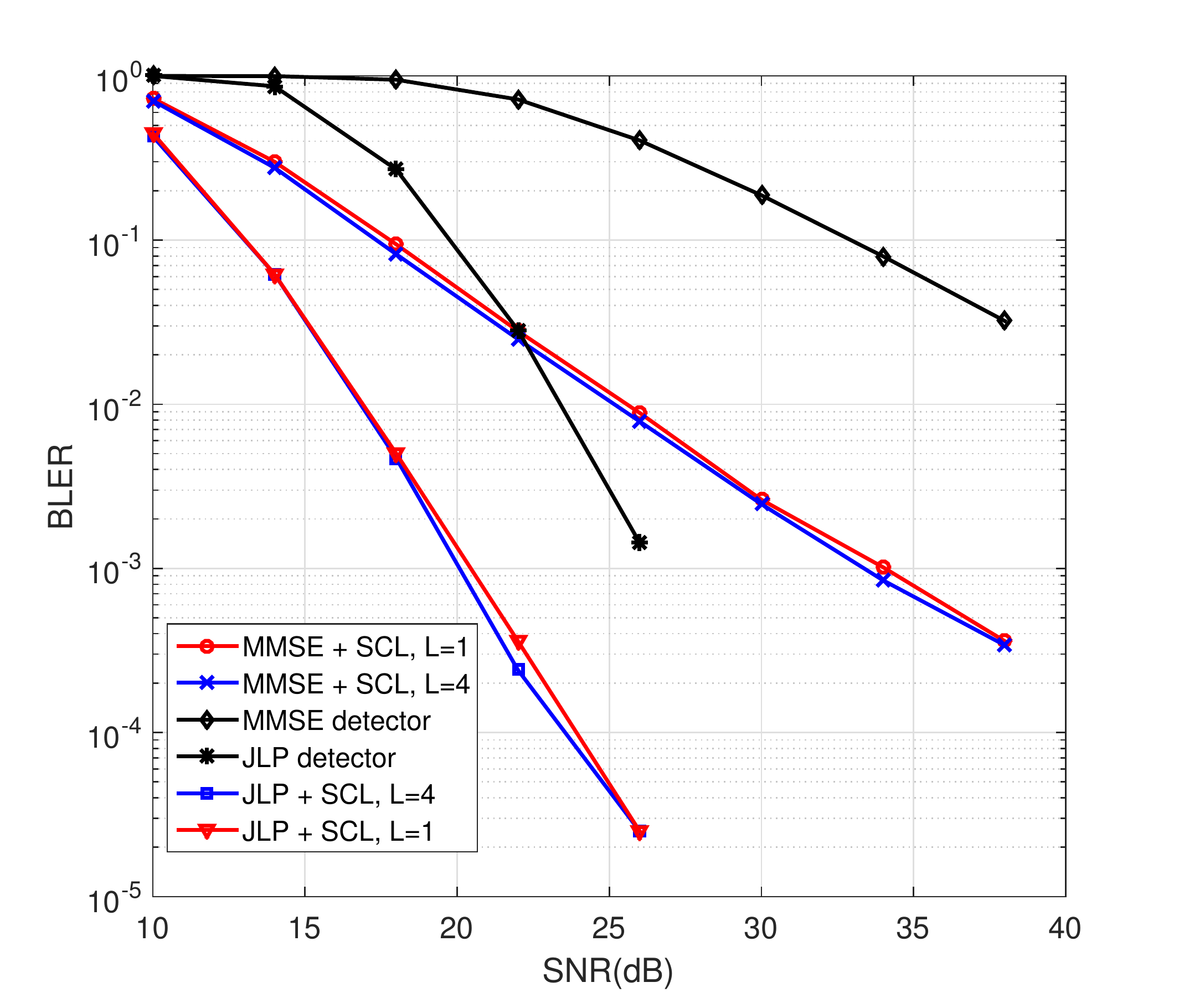}
    \includegraphics[scale = 0.4]{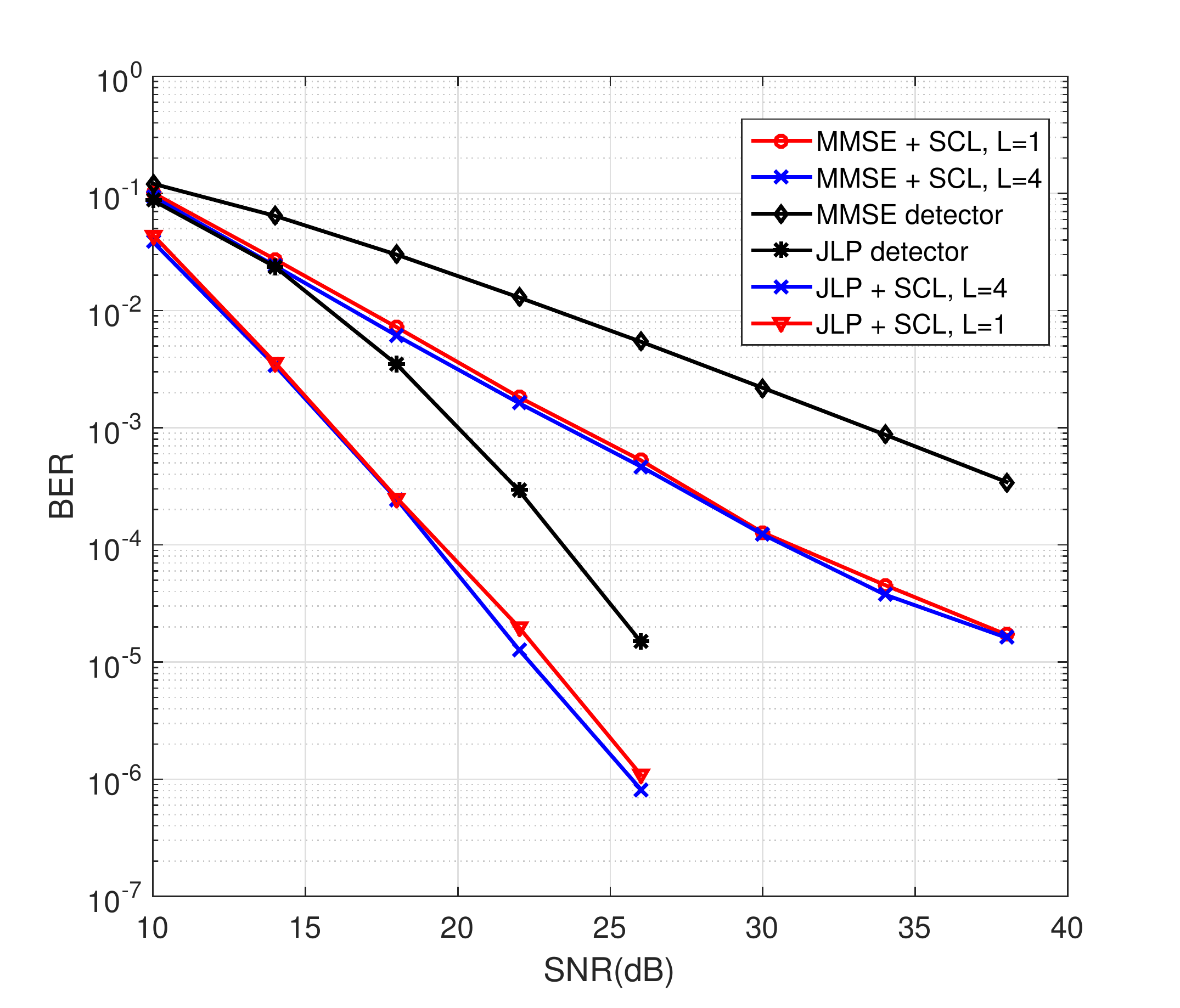}
    \caption{Performance comparison of joint LP detector using code constraints versus decoupled MMSE, $\alpha = 2$, $4 \times 4$ MIMO}
     \label{fig:fig5}
\end{center}
\end{figure} 

Figure~\ref{fig:fig3} shows the BER and BLER of the proposed JLP receiver with the benchmark MMSE receiver for code length of
$N=128$. Three different parameter settings are considered, respectively. 
The first set of results compares the JLP detector without SCL decoding against
the MMSE detector without SCL decoding. It can be clearly seen from 
Fig.~\ref{fig:fig3} that our proposed JLP receiver substantially outperforms 
the MMSE detector by integrating the FEC codeword constraint information
during detection.  Thus, this detector output that incorporates the polar code
constraints can further improve the BER by 
as much as 12dB in terms of signal-to-noise ratio (SNR) at the BER of $10^{-3}$. 
We further compare the effect of SCL decoding after detection output
based on JLP against MMSE detection.  It is clear from Fig.~\ref{fig:fig3} that
both BLER and BER are substantially reduced when ensuing polar decoding
is adopted. In terms of BLER,  the performance improvement by the JLP over
MMSE can be as high as 10dB for both SCL based on $L=1$ and $L=4$.


Fig. ~\ref{fig:fig5} illustrates the performance comparison when channel estimation error variance is 
quite substantial, at twice the channel white noise variance, i.e. $\alpha = 2$. In this 
case the performance gain of JLP over MMSE is even more pronounced. 
JLP together with SCL decoder outperforms MMSE with SCL decoder by nearly $16$ dB at the
BER level of $1.5 \times 10^{-5}$.
The performance improvement in JLP is to be expected since MMSE only relies 
the received signals without integrating the code constraints. 
Thus, MMSE signal detection can be more vulnerable to channel state information (CSI)
error and would have no weapon to fight against CSI errors.  
On the other hand, JLP takes advantage of FEC coding constraints 
and can be more robust against both CSI error and additive channel
noise.

\begin{figure}
 \begin{center}
    \includegraphics[scale = 0.43]{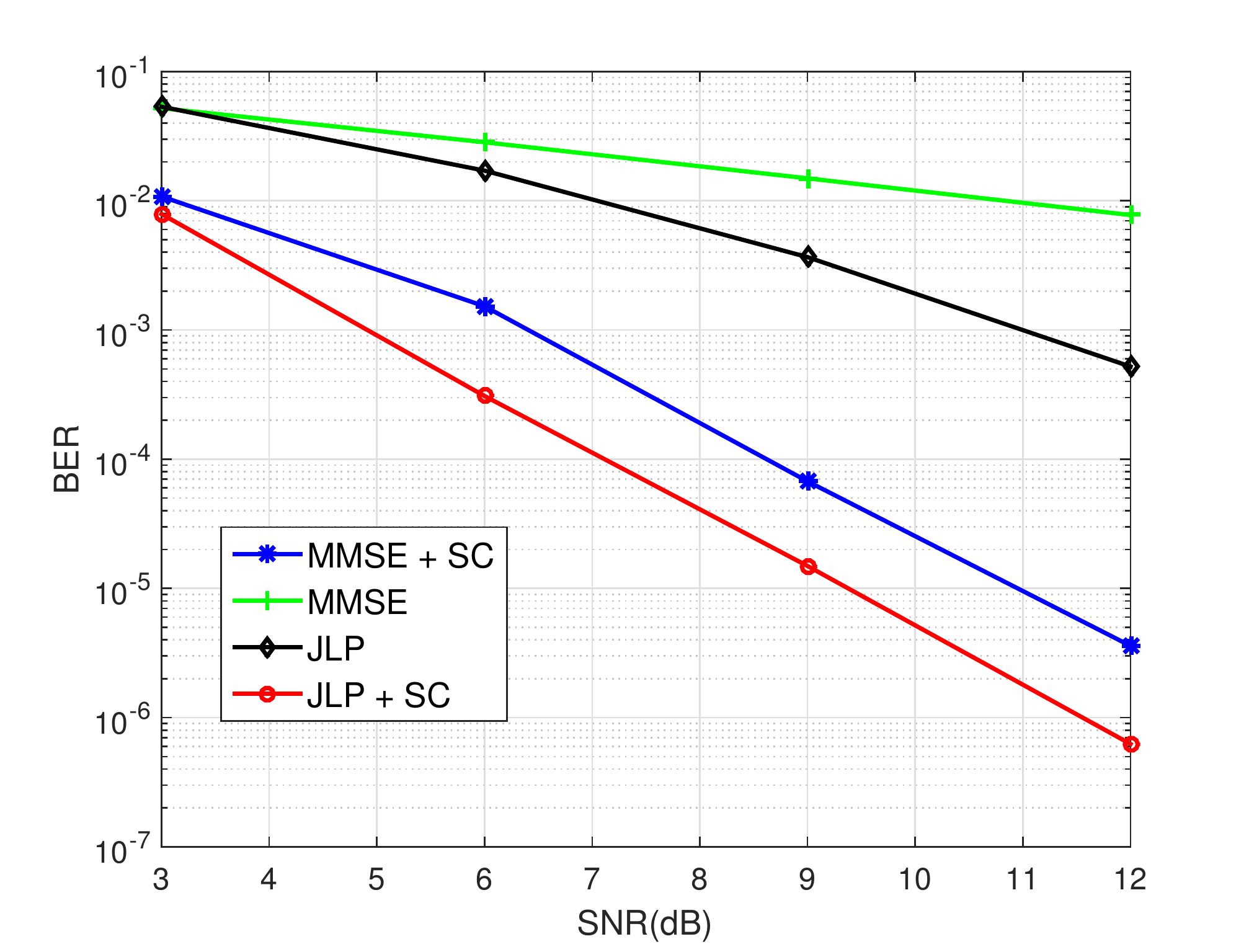}
    \includegraphics[scale = 0.49]{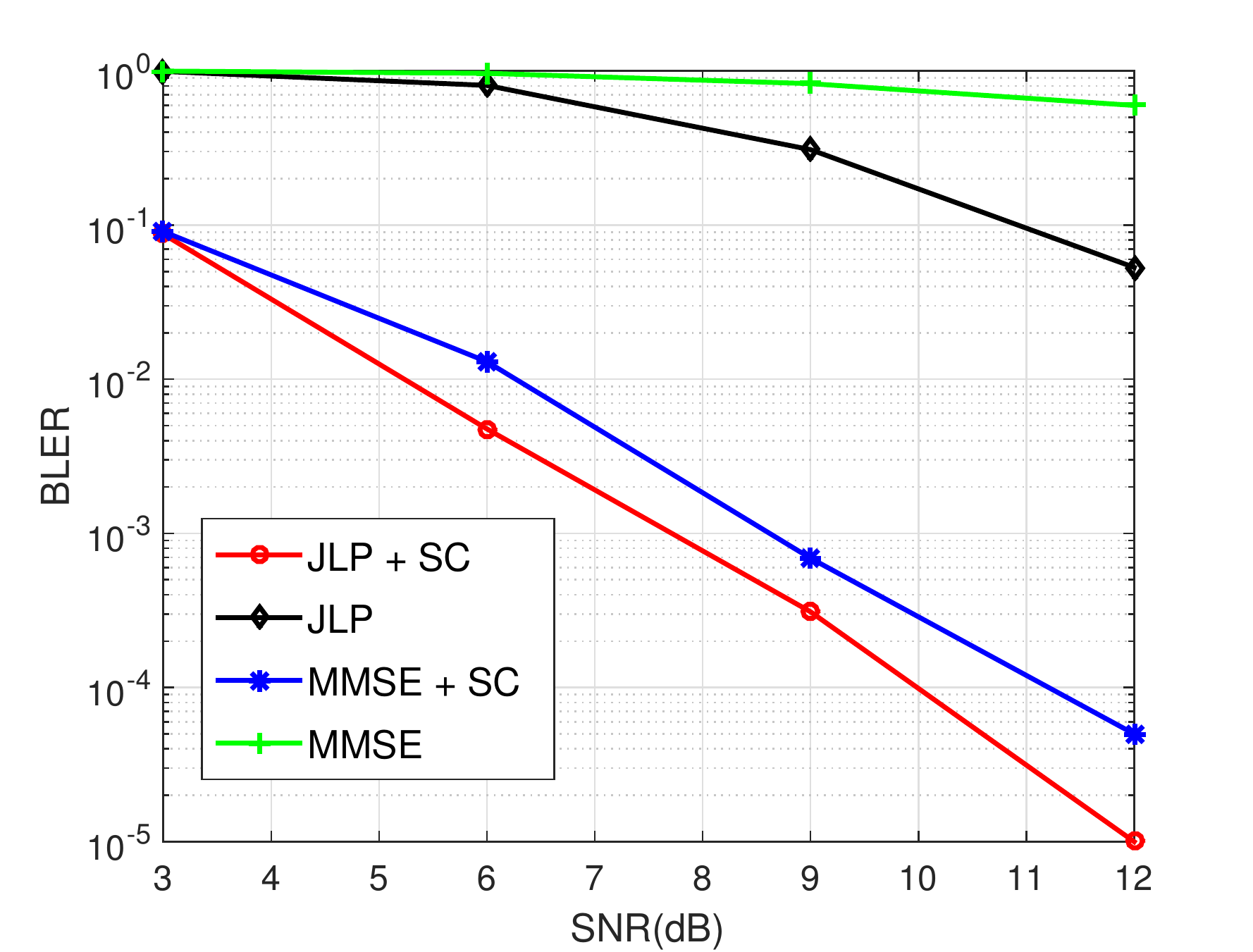}
    \caption{Performance comparison of joint LP detector using code constraints versus decoupled MMSE, $\alpha = 1$, $4 \times 1$ MIMO}
     \label{fig:fig6}
\end{center}
\vspace*{-2mm}
\end{figure} 

In Figure~\ref{fig:fig6}, we also test the reliability of the two detectors 
in case of a $4 \times 1$ transmit diversity MIMO transmission. It is obvious
that the channel uncertainty becomes less of an impairment than in the $4 \times 4$ case,
as the $4\times 1$ transmission diversity provides stronger resilience. 
Nevertheless, the JLP still outperforms the MMSE by approximately 1.5 dB, 
although the gap is less compared to $4 \times 4$ case.

\subsection{Blind Detection of PDCCH}
We now present the simulation results for a blind detection scenario to demonstrate the effectiveness of 
our proposed scheme. LP detector receives $C=20$ PDCCH candidates with length $N = 128$ plus
polar code rate of $0.5$. Only one of candidates has the DCI relevant to the UE, 
as a result, after UE demasks its tested $C$ candidates with its own UE-specific RNTI, 
at most one candidate may have the structure of a valid polar code while the remaining $C-1$ 
DCI candidates would be equivalent to random blocks of bits.

We ran a Monte Carlo simulation with $10^6$ trials. The channel is assumed to be a $4\times 1$
transmit diversity MIMO channel. Both perfect and imperfect CSIs are studied. 
We measure the performance of our blind detection scheme in terms of decoder usage reduction as well
as the missed detection rate. 
Decoder usage reduction measures the ratio of instances in our simulations that our scheme is able to correctly 
identify the right candidate at the detector and therefore decoder usage will be only $1$ instead of $20$. 
Missed detection is defined as the probability that our blind detection scheme fails to 
find its own candidate and, therefore, the decoder fails to decode the right DCI, leading to
lost payload. 

In Table \ref{table:1} and Table \ref{table:2},
it is clear that our newly proposed PDCCH blind detector leads to  
substantial saving of decoder usage at various values of SNR and $\beta$
for both perfect and imperfect CSI, respectively.

 \begin{table}
  \centering
\caption {Decoder usage reduction for perfect channel estimation}
\label{table:1}
 \begin{tabular}{ |p{1cm}|p{1cm}|p{2cm}|p{2cm}|  }
\hline
SNR& $\beta$ &missed detection & decoder usage reduction\\
 \hline
 
 3   &   4     & $3 \times 10^{-5}$   &   51.56\%\\
 3   &   4.5  & $8 \times 10^{-6}$   &   38.00\%\\
 3   &   5     & $2 \times 10^{-6}$   &   26.10\%\\
 3   &   5.5     & 0                           &   16.83\%\\
  \hline
 5   &   4     & $3 \times 10^{-6}$   &   78.80\%\\
 5   &   4.5  & 0                              &   66.62\%\\
 5   &   5     & 0                              &   52.26\%\\
  \hline
 8   &   4     & 0                              &   97.51\%\\
 8   &   4.5  & 0                              &   93.70\%\\
 8   &   5     & 0                              &   86.42\%\\
 \hline

\end{tabular}
\end{table}

 \begin{table}
  \centering
\caption {Decoder usage reduction for imperfect channel estimation}
\label{table:2}
 \begin{tabular}{ |p{1cm}|p{1cm}|p{2cm}|p{2cm}|  }
 \hline
SNR& $\beta$ &missed detection & decoder usage reduction\\
 \hline
 8   &   4     & $2.2 \times 10^{-5}$   &   56.21\%\\
 8   &   4.5  & $3 \times 10^{-6}$      &   41.73\%\\
 8   &   5     & $10^{-6}$      &   28.76\%\\
 8  &   5.5     & 0                               &   18.51\%\\
 \hline
11   &   4     &  0  		               &   90.71\%\\
11   &   4.5  &  0 			       &   81.83\%\\
 11  &   5     &  0  			       &   69.26\%\\

 \hline

\end{tabular}
\end{table}

Recall that loss of DCI will cause the loss of DL-SCH payload.  As a result, 
the gNB has to retransmit the DCI and DL-SCH again. 
Thus, missed detection of DCI would severely increase latency and deduce the throughput of the wireless system. 
Therefore, we need to set $\beta$ such that missed detection probability is as low as possible at the risk
of increased decoder usage. 
From the results in Table~\ref{table:1} and Table~\ref{table:2} above,
$\beta \geq 5.5$ leads to a missed detection probability
that is low enough not to even show up in our Monte Carlo simulation of $10^6$ trials.
Yet, we can still benefit from the significant decoder usage reduction in moderate SNR scenarios. 
For instance, in case of perfect CSI, when SNR$\,= 8$ and $\beta = 5$, 
our receiver can identify the right DCI candidate $86.42\%$ of the time while only running the SCL decoder once 
instead of $20$ times. In case of imperfect CSI, 
higher SNR is needed to achieve the same performance. 
When SNR $= 11$ and $\beta = 5$, the decoder usage reduction of $69.26\%$ is achieved. 

Our simulation results demonstrate the benefit of our new receiver architecture. 
Both missed detection and decoder use reduction have been substantially contained.

\section{Conclusion}

This work considers the important and practical problem of blind detection for user DCI in PDCCH
of 5G New Radio cellular systems. To jointly reduce the receiver complexity and the probability
of missed detection,  we proposed a robust joint detection-decoding receiver design in
MIMO systems that adopt the effective  polar codes for forward error correction (FEC). 
We incorporate relaxed polar code constraints to formulate a novel joint LP optimization problem. 
The proposed receiver is more robust against CSI errors, channel noises, and 
other non-idealities at the wireless receivers.
We also introduced a metric at the LP detector that is capable of identifying the right PDCCH candidate 
while rejecting the false ones. The use of this metric improves the so called blind detection process in NR standard.
Our proposed joint LP detector can also 
directly interface with well-known polar decoders such as SC and SCL 
decoders for effective error correction. As a result, JLP detector can be an appropriate receiver candidate to be used for control channels of 5G systems. 
Our test results demonstrate superior receiver performance particularly 
when CSI knowledge s inaccurate due to practical obstacles such
as short pilot length and pilot contamination. 
The results further show that our proposed
fractional metric can identify the right candidate in moderate to high SNR scenarios with negligible probability of missed detection.

\ifCLASSOPTIONcaptionsoff
  \newpage
\fi



\bibliographystyle{IEEEtran}
\bibliography{reference}
\end{document}